\documentstyle[12pt,epsfig,rotating]{article}

\newcommand{\mysection}{\setcounter{equation}{0}\section}

\newcommand{\gve}{{g^V_e}}
\newcommand{\gae}{{g^A_e}}
\newcommand{\gvq}{{g^V_q}}
\newcommand{\gaq}{{g^A_q}}
\newcommand{\lama}{\lambda_{e^+}}
\newcommand{\lamb}{\lambda_{e^-}}
\newcommand{\Slash} {\slash \!\!\!}
\newcommand{\Li}{{\cal L}i_2}
\newcommand{\sr}{\sqrt{\rho}}
\newcommand{\cost}{\cos \theta }
\newcommand{\costs}{\cos^2 \theta}

\newcommand{\sints}{\sin^2 \theta}
\newcommand{\wnw}{W_T}
\newcommand{\nw}{\hat n \cdot \hat W}
\newcommand{\sqt}{\sqrt{t}}
\newcommand{\den}{(4 -4 x +\rho)}

\newcommand{\pgg}{{\cal P}^{\gamma \gamma}}
\newcommand{\pzz}{{\cal P}^{Z Z}}
\newcommand{\Rgz}{{\rm Re}{\cal P}^{\gamma Z}}
\newcommand{\Igz}{{\rm Im}{\cal P}^{\gamma Z}}
\begin{document}
\begin{flushright}
MRI-O-000601\\
\end{flushright}
\begin{flushright}
INLO-PUB-11/2000
\end{flushright}
\begin{flushleft}
{\tt hep-ph/0006125}\\
\end{flushleft}
\vskip 1.0cm
\centerline{\large\bf { QCD corrections up to order $\alpha_s^2$ to 
polarized quark }}
\centerline{\large\bf {production in $e^+e^-$ -annihilation}}
\vskip 2.0cm
\centerline {\sc V. Ravindran}
\centerline{\it Mehta Research Institute of Mathematics and Mathematical
Physics,}
\centerline{\it Chhatnag Road, Jhusi, Allahabad-211019,}
\centerline{\it India.}
\vskip 1.0cm
\centerline {\sc W.L. van Neerven}
\centerline{\it Instituut-Lorentz,}
\centerline{\it University of Leiden,}
\centerline{\it PO Box 9506, 2300 RA Leiden,}
\centerline{\it The Netherlands.}
\vskip 1.0cm
\centerline{June 2000}
\vskip 2.0cm
\centerline{\bf Abstract}
\vskip 0.3cm
We present the calculation of the order $\alpha_s^2$ contributions to the
cross section $e^+ + e^- \rightarrow \bar q + q$ where the incoming leptons as
well as one of the outgoing (anti) quarks are longitudinally polarized. 
The computation is carried out for massless quarks so that it can be applied 
to light flavour production ($u, d, s$). Unfortunately the massless quark
approach is not valid for heavy flavour production like $c, b, t$ even in 
the case when the centre of mass energy $Q$ is much larger than the quark mass
$m$. This is in contrast to unpolarized scattering where this approach
works rather well for $Q\gg m$. The reason for this can be attributed to the 
anomalous terms which are characteristic of polarized coefficient functions.
Furthermore we also computed the order $\alpha_s$ corrections to the
longitudinal, transverse and normal polarizations of heavy flavours with 
$m \not =0$.
The latter have been presented earlier in the literature except for some
contributions which are shown here for the first time.
It turns out that the corrections to the longitudinal and transverse (in the
plane) polarization are rather small. However the order $\alpha_s$ corrections
to the normal (out of the plane) polarization are large so that second
order contributions (for $m \not =0$) are needed to get a better determination 
of this quantity.\\[3mm]

\mysection{Introduction}
Quark production in electron positron annihilation provides us with
additional information about the constants appearing in the standard model
of the electroweak and strong interactions. This in particular holds
for heavy flavour production. An example is the electroweak mixing angle
$\theta_W$ which has been very accurately extracted from the forward-backward
asymmetry measured for bottom quarks at LEP and SLC. Besides a more accurate 
determination of these constants heavy quark production can also reveal 
some signatures of new physics. They might be observable when linear colliders 
are built \cite{zerw} which are giving access to much larger energies
than those available until now. These energies are large enough
to observe top anti-top production which will be one of the issues
under study. In particular of interest is the polarization of this quark
because the spin can be easily measured from its decay products. This is 
possible because the mass of the top-quark is so big that it can decay 
electroweakly before it undergoes hadronization. The measurement of the 
polarization provides us with new tests of the standard model and it
might even signal new interactions which are not predicted by this model.
An example is given in \cite{ritu} where the effect of anomalous chromoelectric
couplings of the gluon to the top quark on the longitudinal polarization
is studied. Another important quantity is the polarization which is 
perpendicular to the plane spanned by the momenta of the incoming electron
and outgoing quark. This so called normal polarization is a time reversal
odd observable which has implications for the observation of CP-violation in 
the neutral current sector. Polarized heavy quark production has been studied 
on the Born level in \cite{craig}, \cite{krz}, \cite{akp}. In most of these
calculations the top quark spin is decomposed in the helicity basis.
However one can also make other choices like the beamline basis and
the off-diagonal basis. In \cite{pash} one has shown that choosing
the off-diagonal basis the top and anti-top quarks are produced in one unique
spin configuration only which depends on the helicities of the incoming leptons.
QCD corrections to quark production have been computed in several papers 
\cite{krz}, \cite{stol}-\cite{konapa}. The previous references deal with
the production of heavy quarks only. For the decay process, among which the
spin density matrix, see \cite{schm}, \cite{brfl}. The corrections in the case 
of massive quarks are rather complicated even in first order.
In the case of the helicity basis the most of them are done by one group only 
see \cite{kopitu}-\cite{grkotu2}
except for a few corrections which will be presented in this paper. 
The QCD corrections in the case of the off-diagonal basis have been calculated 
in \cite{konapa} as far as the soft and virtual gluon corrections are 
concerned. However the contribution due to hard gluon bremsstrahlung has not 
been calculated in this basis. Therefore the authors in \cite{konapa} choose
an alternative by putting one of the quarks in a special spin configuration 
whereas the other (anti-) quark and the gluon were taken
to be inclusive. Furthermore one has neglected the width of the Z-boson 
and the normal polarization was not considered. Because of the 
importance of polarized quark production one should have an independent check 
on the calculations in the literature. Therefore we want to repeat the 
computations done in \cite{kopitu}-\cite{grkotu2} and include some
order $\alpha_s$ corrections which were not considered before. Because of
the complexity of the expressions for the first order corrections
we try to write the results as compact as possible. The procedure is akin to 
the one followed in \cite{rane1} in which the forward-backward asymmetry and 
the shape 
parameter could be written in a compact form. The same was also achieved
for the longitudinal and transverse cross section in \cite{rane2}. 
The main problem is to reduce the number of Spence functions because they lead
to unnecessarily long expressions. Exploiting
several relations like the Hill-identity \cite{lbmr} one can reduce them to a 
minimum. Furthermore we will present the cross section in such a way that it
holds for the longitudinal, transverse and normal polarization
at the same time. Finally we concentrate on the second order corrections to 
the longitudinal polarization of massless quarks. They can be derived from the 
first moment of the timelike coefficient functions computed in 
\cite{rine1}-\cite{rine2}. In the case of unpolarized scattering \cite{rane1}, 
\cite{rane2} the massless quark approach can be also applied to
heavy quarks provided the centre of mass (CM) energy is much larger than the 
mass of the quark. These second order estimates are very useful 
\cite{case} as long as the exact calculations are not available because the 
latter are very difficult to compute. 
Unfortunately the massless quark approach does not work for heavy flavour
production in the case of polarized scattering. In this approach the
coefficient functions are computed under the condition that the quark mass
m is put to zero at the start of the calculation like in \cite{rane1},
\cite{rane2} (for the first order see \cite{flsa}, \cite{ra}, \cite{stvo}).
One can also compute these functions in the so called massive quark 
approach with $m \not =0$ and taking the limit $m \rightarrow 0$ afterwards. 
In the calculation of the polarized coefficient functions both approaches 
lead to different results contrary to what we have seen for unpolarized
scattering.
In other words the zero mass limit does not commute with the integrations.
This anomaly is due to chiral symmetry breaking when the quark becomes massive 
and it was discovered for the first Bjorken sum rule which is given
by the first moment of the longitudinal spin structure function $g_1(x,Q^2)$
in \cite{mene}, \cite{teve}. For timelike processes like $e^+e^-$ collisions 
it was discussed in \cite{grkotu1}, \cite{grkotu2}, \cite{fase}. Here we would 
like to emphasize that this anomaly does not affect the longitudinal 
polarization of the light flavours even if we regularize the collinear
divergences by giving the light quark a fictitious mass. It turns out
that this anomaly also appears in the quark operator matrix element so that 
it will be removed by mass factorization (see \cite{mene}). Hence we obtain the
same result as for $m=0$ where one can use n-dimensional regularization. 
However for heavy flavours, where the mass has a definite meaning, a 
subtraction via the quark operator matrix element is not justified so that
these anomalous terms are retained in the QCD corrections. 
Therefore in the case of heavy quarks the polarized coefficient functions have 
to be calculated for non zero masses. These calculations are far from trivial
because one cannot apply the tricks which were so successful for the
calculation in \cite{chkust} of the order $\alpha_s^2$ corrections to 
$\sigma_{tot}(e^+~e^- \rightarrow `hadrons')$ with massive quarks. 
This is because the timelike coefficient functions cannot be written as
the imaginary part of a forward scattering amplitude which is an essential
ingredient for the computations in \cite{chkust}.
The paper will be organized as follows. In the section 2 we define the 
kinematics and give an outline of the calculation procedure. We also
present the formulae for the spin dependent and spin independent parts of
the cross section. In section 3 we show the results for the radiative
corrections computed up to first order in the strong coupling constant
$\alpha_s$ for massive quarks. Here we also include the contributions
which are proportional to the width of the Z-boson. The corrections will
be extended for the longitudinal polarization of massless quarks up to order 
$\alpha_s^2$. In section 4 we discuss the effects of these corrections on the 
longitudinal, transverse and normal polarization of the detected quark.
The long expressions for the order $\alpha_s$ corrected quark structure 
functions $W_i(x,Q^2,m^2)$ are presented in Appendix A. After integrating
these functions over the Bjorken variable $x$ we could express them into a 
compact form in Appendix B.


\mysection{Kinematics for polarized $e^+~e^-$-annihilation}
Polarized heavy quark production in electron-positron annihilation is given 
by the following process
\begin{eqnarray}
\label{eqn:2.1}
e^+(q_{e^+},\lama) + e^-(q_{e^-},\lamb) \rightarrow 
V(q) \rightarrow {\rm H}(p,s) + "{\rm X}" \,,
\end{eqnarray}
where $\rm H$ denotes the heavy quark and $"{\rm X}"$ represents any inclusive 
multi-partonic state containing the heavy anti-quark and the light (anti-)
quarks and gluons.
Further $\lambda_i,q_i$ denote the spin and momenta of the
incoming leptons and $s,p$ stand for the spin and momentum of the heavy quark
$\rm H$ detected in the final state. If we denote the momentum of the 
virtual vector boson $V$ ($V=\gamma, Z$) by $q$, the
centre of mass energy $Q$ is defined by
\begin{eqnarray}
\label{eqn:2.2}
q^2=Q^2=(q_{e^+}+q_{e^-})^2\,.
\end{eqnarray}
The differential cross section corresponding to reaction (\ref{eqn:2.1}) equals
\begin{eqnarray}
\label{eqn:2.3}
d \sigma &=& {1 \over 2 Q^2} dPS~ \sum_{V_1V_2} {\cal L}_{\mu \nu}^{(V_1V_2)}
{\cal P}^{(V_1V_2)} ~M^{\mu \nu,(V_1V_2)} 
\nonumber\\[2ex]
& \equiv & {1 \over 2 Q^2} dPS~ \left [ {\cal L}_{\mu \nu}^{(\gamma \gamma)}~
{\cal P}^{(\gamma \gamma)} ~M^{\mu \nu,(\gamma \gamma)} \right.
\left. + {\cal L}_{\mu \nu}^{(ZZ)}~{\cal P}^{(ZZ)} ~M^{\mu \nu,(ZZ)} \right.
\nonumber\\[2ex]
&& \left. +{\cal L}_{\mu \nu}^{(\gamma Z)}~{\cal P}^{(\gamma Z)}
~M^{\mu \nu,(\gamma Z)} 
+{\cal L}_{\mu \nu}^{(\gamma Z)*}~{\cal P}^{(\gamma Z)*}~
M^{\mu \nu,(\gamma Z)*} \right ]\,.
\end{eqnarray}
Here $dPS$ denotes the multi-parton phase space including the heavy quark and 
${\cal L}_{\mu \nu}^{(V_1V_2)}$ and $M_{\mu \nu}^{(V_1V_2)}$ are the 
leptonic and partonic matrix elements respectively. Further 
${\cal P}^{(V_1V_2)}$ 
represents the product of the propagators corresponding to the vector bosons 
$V_1$ and $V_2$. The leptonic tensors are given by
\begin{eqnarray}
\label{eqn:2.4}
{\cal L}_{\mu \nu}^{(\gamma\gamma)}&=& 4 \pi \alpha ~ Q_e^2
                 \Bigg [~ (1-\lama \lamb)~ l_{\mu \nu}
                 + (\lama-\lamb)~ \tilde l_{\mu \nu} ~\Bigg] \,,
\nonumber\\[2ex]
{\cal L}_{\mu \nu}^{(ZZ)}&=& {4 \pi \alpha \over c_w^2 s_w^2} ~
                  \Bigg[~ \left(-2 \gve \gae (\lama-\lamb)
                         +(\gve^2+\gae^2)(1-\lama \lamb) \right)~ l_{\mu \nu} 
\nonumber\\[2ex]
&&                 + \left(-2 \gve \gae (1-\lama \lamb)
                         +(\gve^2+\gae^2)(\lama-\lamb)\right)~
                 \tilde l_{\mu \nu}  ~\Bigg] \,,
\nonumber\\[2ex]
{\cal L}_{\mu \nu}^{(\gamma Z)}&=& - {4 \pi \alpha \over c_w s_w} Q_e ~
                  \Bigg[~ \left(- \gve (1-\lama \lamb)
                         +\gae (\lama- \lamb) \right)~ l_{\mu \nu} 
\nonumber\\[2ex]
&&                 + \left(- \gve (\lama- \lamb)
                         +\gae(1-\lama \lamb)\right)~
                 \tilde l_{\mu \nu}  ~\Bigg]\,,
\end{eqnarray}
with
\begin{eqnarray}
\label{eqn:2.5}
l^{\mu \nu}&=& q^{\mu}_{e^+} q^{\nu}_{e^-}+ q^{\nu}_{e^+} q^{\mu}_{e^-} 
-q_{e^+}.q_{e^-} g^{\mu \nu} \,,
\nonumber\\[2ex]
\tilde l^{\mu \nu}&=&i \epsilon^{\mu \nu \alpha \beta} q_{e^+\alpha} 
q_{e^-\beta}\,.
\end{eqnarray}
Notice that the polarizations of the positron and the electron indicated by
$\lama$ and $\lamb$ respectively are defined by 
\begin{eqnarray}
\label{eqn:2.6}
v_{\lama}(q_{e^+}) \bar v_{\lama}(q_{e^+}) &=& {1 \over 2}
\Slash q_{e^+}(1+\lama \gamma_5) \,,
\nonumber\\[2ex]
u_{\lamb}(q_{e^-}) \bar u_{\lamb}(q_{e^-}) &=& {1 \over 2}
\Slash q_{e^-}(1-\lamb \gamma_5)\,,
\end{eqnarray}
where the incoming leptons are taken to be massless. In the case $\lambda_i=1$
the leptons are polarized along the direction of their momenta. When 
$\lambda_i=-1$ the leptons are polarized opposite to the direction of their 
momenta.

On the Born level the vertex for the coupling of the vector boson $V$ 
to the fermions will be denoted by
\begin{eqnarray}
\label{eqn:2.7}
\Gamma_{a,\mu}^{V,(0)}&=&i~({\it v}_a^V + {\it a}_a^V \gamma_5) \gamma_\mu\,,
\qquad V=\gamma, Z\,,
\nonumber\\[2ex]
{\it v}_a^{\gamma}&=& -e Q_a \,, \qquad {\it a}_a^{\gamma}= 0\,,
\nonumber\\[2ex]
{\it v}_a^Z & =& - { e \over c_w s_w} g^V_a\,, \qquad {\it a}_a^Z  =
{e \over c_w s_w} g^A_a\,.
\end{eqnarray}
The electroweak coupling constants are given by
\begin{eqnarray}
\label{eqn:2.8}
\alpha&=&e^2/4 \pi\,,\quad c_w=\cos \theta_W\,,\quad s_w=\sin \theta_W\,,
\nonumber\\[2ex]
g^V_a&=&{1 \over 2}  T^3_a -s_w^2 Q_a\,,
\qquad g^A_a=-{1 \over 2} T^3_a\,.
\end{eqnarray}
The electroweak charges for the leptons are equal to
\begin{eqnarray}
\label{eqn:2.9}
Q_a&=&0\,, \quad T^3_a=\frac{1}{2}\,, \quad a=\nu_l, \bar \nu_l \,,
\quad l=e, \mu, \tau \,,
\nonumber\\[2ex]
Q_a&=&-1\,, \quad T^3_a=-\frac{1}{2} \quad a=e^-,\mu^-,\tau^-\,,
\end{eqnarray}
and for the quarks we obtain
\begin{eqnarray}
\label{eqn:2.10}
Q_a&=&\frac{2}{3}\,, \quad T_3^a=\frac{1}{2} \,, \quad a=u,c,t\,,
\nonumber\\[2ex]
Q_a&=&-\frac{1}{3}\,, \quad T_3^a=-\frac{1}{2} \,, \quad a=d,s,b\,.
\end{eqnarray}
The squared propagators ${\cal P}^{(V_1V_2)}$ are given by
\begin{eqnarray}
\label{eqn:2.11}
{\cal P}^{\gamma\gamma} &=& { 1\over Q^4}\,, \qquad {\cal P}^{ZZ} 
= { 1\over (Q^2-M_Z^2)^2+M_Z^2\Gamma_Z^2}\,,
\nonumber\\[2ex]
 {\cal P}^{\gamma Z} &=& 
{ (Q^2-M_Z^2) + i~M_Z \Gamma_Z \over Q^2 \{ (Q^2-M_Z^2)^2 + M_Z^2\Gamma_Z^2\} }
\,,
\end{eqnarray}
where we have introduced a finite width $\Gamma_Z$ for the Z-boson.\\
The differential cross section for the inclusive reaction Eq. (\ref{eqn:2.1})
can be written as
\begin{eqnarray}
\label{eqn:2.12}
\frac{d \sigma}{d\Omega}&=& {1 \over 2 Q^2} \int_{\sr}^1 dx ~ 
\sum_{(V_1V_2)}~ 
          {\cal L}_{\mu \nu}^{(V_1V_2)} ~{\cal P}^{(V_1V_2)}~
          W^{\mu \nu,(V_1V_2)} \,,
\nonumber\\[2ex]
d~\Omega &=& d\cos \theta~ d \phi \,, \qquad x = \frac{2p \cdot q}{Q^2}\,,
 \qquad \rho= \frac{4m^2}{Q^2}\,.
\end{eqnarray}
Here $\theta$ is the polar angle of the outgoing quark with respect to the 
beam direction of the electron and $m$ denotes the mass of the heavy quark.
The partonic tensor can be expressed into structure functions 
$W_i^{(V_1V_2)}$ ($i=1-11$) in the following way
\begin{eqnarray}
\label{eqn:2.13}
W^{\mu \nu,(V_1V_2)}&=& {N_c \over 8 \pi^2 } Q^2 \Bigg(
   g^{\mu \nu} W_1^{(V_1V_2)}+{p^{\mu} p^{\nu} \over Q^2}  W_2^{(V_1V_2)}
 + m g^{\mu \nu} {s \cdot q \over Q^2} W_3^{(V_1V_2)}
\nonumber\\[2ex]
&& +m p^{\mu} p^{\nu} {s \cdot q \over Q^4} W_4^{(V_1V_2)}
     +m {p^{\mu} s^{\nu} +s^{\mu} p^{\nu} \over Q^2}  W_5^{(V_1V_2)}
\nonumber\\[2ex]
&& +{1 \over Q^2} i \epsilon^{\mu \nu \alpha \beta} p_{\alpha} q_\beta 
W_6^{(V_1V_2)} +m i \epsilon^{\mu \nu \alpha \beta} p_{ \alpha} q_\beta  
{s \cdot q \over Q^4} W_7^{(V_1V_2)}
\nonumber\\[2ex]
&& + {m \over Q^2}i \epsilon^{\mu \nu \alpha \beta} p_{\alpha} s_\beta  
W_8^{(V_1V_2)}
 + {m\over Q^2} i \epsilon^{\mu \nu \alpha \beta} q_{\alpha} s_\beta
W_9^{(V_1V_2)} 
\nonumber\\[2ex]
&&     +m {p^{\mu} s^{\nu} -s^{\mu} p^{\nu} \over Q^2}  W_{10}^{(V_1V_2)}
    + {m\over Q^4} i \Big(p^\mu\epsilon^{\nu \alpha \beta \gamma} 
p_{\alpha} q_\beta s_\gamma
\nonumber\\[2ex]
&&     +p^\nu  \epsilon^{\mu \alpha \beta \gamma} p_{\alpha} q_\beta
     s_\gamma \Big) W_{11}^{(V_1V_2)} \Bigg)\,,
\end{eqnarray}
where $N_c$ denotes the number of colours (in QCD $N_c=3$).
Finally we have to specify the spin of the quark ${\rm H}(p,s)$.
In the rest frame i.e. $p=(m,\vec 0)$ the spin vector is given by
\begin{eqnarray}
\label{eqn:2.14}
s=(0,\hat W)\,, \quad \hat W=(\hat W^1,\hat W^2,\hat W^3) \,, \quad
\mbox{with} \quad \hat W^2=1\,,
\end{eqnarray}
so that $s^2=-1$ and $s\cdot p=0$.
In the CM frame of the electron positron pair we introduce the notations
\begin{eqnarray}
\label{eqn:2.15}
&& q_{e^-}=\frac{Q}{2}(1,0,0,1)\,, \qquad q_{e^+}=\frac{Q}{2}(1,0,0,-1)\,,
\nonumber\\[2ex]
&& p=\left (E,|\vec p| \hat n \right) \qquad 
n=\left (\sin \theta \cos \phi, \sin \theta \sin \phi, \cos \theta \right )\,,
\nonumber\\[2ex]
&& E=\frac{1}{2} Q x \,, \qquad |\vec p|=\frac{1}{2} Q \sqrt{ x^2 -\rho}\,.
\end{eqnarray}
In this frame The spin four-vector of the quark can be found after an 
appropriate Lorentz transformation so that it becomes equal to
\begin{eqnarray}
\label{eqn:2.16}
s=\left(|\vec p| {\hat n\cdot \hat W \over m}, \hat W
    +{|\vec p|^2 (\hat n \cdot \hat W ) \hat n \over m(m+E)} \right) \,,
\quad \mbox{with} \quad \hat n^2=1\,.
\end{eqnarray}
Note that the spin four-vector in the CM frame
depends on the energy of the quark and hence depends on the integration 
variable $x$ in Eq. (\ref{eqn:2.12}). 
The computation of the cross section in Eq.
(\ref{eqn:2.12}) involves the contraction of the symmetric and
antisymmetric parts of the leptonic tensor ${\cal L}_{\mu \nu}^{(V_1V_2)}$
with $W^{\mu \nu,(V_1V_2)}$. The contraction with the symmetric part equals
\begin{eqnarray}
\label{eqn:2.17}
\!\!\!\int dx~ l_{\mu \nu}~ W^{\mu \nu,(V_1V_2)}\!\!\! &=& \!\!\!
{N_c Q^4 \over 8 \pi^2} \Bigg[ {1 \over 8} (1+ \cos^2 \theta)
 \left({\it v}_q^{V_1}{\it v}_q^{V_2} {\cal T}_1 +{\it a}_q^{V_1}
  {\it a}_q^{V_2} {\cal T}_2 \right)
\nonumber\\[2ex]
&& +{1 \over 4} \sin^2 \theta  \left ({\it v}_q^{V_1}{\it v}_q^{V_2}
{\cal T}_3  + {\it a}_q^{V_1} {\it a}_q^{V_2} {\cal T}_4\right)
\nonumber\\[2ex]
&& + { 1 \over 2 }
\left ({\it v}_q^{V_1} {\it a}_q^{V_2} + {\it v}_q^{V_2} {\it a}_q^{V_1}
\right ) {\cal T}_7 
+ { 1 \over 2 }
\left ({\it v}_q^{V_1} {\it a}_q^{V_2} - {\it v}_q^{V_2} {\it a}_q^{V_1}
\right ) {\cal T}_8 
\nonumber\\[2ex]
&& +~{\it v}_q^{V_1} {\it v}_q^{V_2}~ {\cal T}_9 \Bigg]\,,
\end{eqnarray}
and for the antisymmetric parts we get
\begin{eqnarray}
\label{eqn:2.18}
\!\!\!\int dx~ \tilde l_{\mu \nu}~ W^{\mu \nu,(V_1V_2)}\!\!\! &=&\!\!\!
{N_c Q^4 \over 8 \pi^2}
   \Bigg[ {1 \over 4} \left ({\it v}_q^{V_1} {\it a}_q^{V_2}
+ {\it v}_q^{V_2} {\it a}_q^{V_1} \right )
{\cal T}_5 \cos \theta
+{ 1 \over 4} \left ({\it v}_q^{V_1} {\it a}_q^{V_2} \right.
\nonumber\\[2ex]
&& \left. - {\it v}_q^{V_2}{\it a}_q^{V_1} \right ) {\cal T}_6 \cos\theta
+ {\it v}_q^{V_1} {\it v}_q^{V_2} {\cal T}_{10} 
+{\it a}_q^{V_1}{\it a}_q^{V_2} {\cal T}_{11}
\nonumber\\[2ex]
&&+{ 1 \over 2} \left ({\it v}_q^{V_1} {\it a}_q^{V_2} + 
{\it v}_q^{V_2} {\it a}_q^{V_1} \right ) {\cal T}_{12}  
\nonumber\\[2ex]
&& +{ 1 \over 2} \left ({\it v}_q^{V_1} {\it a}_q^{V_2} - {\it v}_q^{V_2} 
{\it a}_q^{V_1} \right ) {\cal T}_{13}  \Bigg]\,,
\end{eqnarray}
In the expressions above ${\cal T}_i$ ($i=1-6$) represent the unpolarized
structure functions.
The polarized structure functions can be decomposed as follows
\begin{eqnarray}
\label{eqn:2.19}
{\cal T}_7&=&{\cal T}_{7,T} \wnw \cost+ W_L \left [{\cal T}_{7,L_1}  ( 1 
+ \costs ) +{\cal T}_{7,L_2} \sints \right ]\,,
\nonumber\\[2ex]
{\cal T}_8&=&{\cal T}_{8,T} \wnw \cost+ {\cal T}_{8,L} W_L  ( 1 + \costs )\,,
\nonumber\\[2ex]
{\cal T}_9&=&{\cal T}_{9,N} W_N   \cos \theta \sin \theta\,,
\nonumber\\[2ex]
{\cal T}_{10}&=&{\cal T}_{10,T} \wnw + {\cal T}_{10,L} W_L \cost\,,
\nonumber\\[2ex]
{\cal T}_{11}&=&{\cal T}_{11,T} \wnw + {\cal T}_{11,L} W_L \cost\,,
\nonumber\\[2ex]
{\cal T}_{12}&=&{\cal T}_{12,N} W_N   \sin \theta\,,
\nonumber\\[2ex]
{\cal T}_{13}&=&{\cal T}_{13,N} W_N    \sin \theta\,.
\end{eqnarray}
Notice that the polarized structure functions ${\cal T}_i$ ($i=7-13$) are 
expanded in terms of the longitudinal ($W_L$), the transverse $W_T$
and the normal polarization $W_N$ of the quark. They are defined by
\begin{eqnarray}
\label{eqn:2.20}
W_L=\nw \,, \qquad
W_T=\hat W^3-\hat n \cdot \hat W \cos \theta \,,
\qquad W_N=\hat W^2 \cos \phi -\hat W^1 \sin \phi\,.
\end{eqnarray}
If we decompose the partonic structure functions as follows
\begin{eqnarray}
\label{eqn:2.21}
W_i^{(V_1V_2)}&=&{\it v}_q^{V_1} {\it v}_q^{V_2} W_i^{{\it v}_q^2}
        +{\it a}_q^{V_1} {\it a}_q^{V_2} W_i^{{\it a}_q^2}
+\left ({\it v}_q^{V_1} {\it a}_q^{V_2}
   +{\it a}_q^{V_1} {\it v}_q^{V_2}\right) W_i^{\{{\it v}_q,{\it a}_q\}} 
\nonumber\\[2ex]
&&      +\left ({\it v}_q^{V_1} {\it a}_q^{V_2}
   -{\it a}_q^{V_1} {\it v}_q^{V_2}\right) W_i^{[{\it v}_q,{\it a}_q]} \,,
\end{eqnarray}
one can express the quantities ${\cal T}_i$ into integrals over the partonic
structure functions $W_i$. For the unpolarized structure functions we obtain
\begin{eqnarray}
\label{eqn:2.22}
{\cal T}_1&=&-4 \int_{\sr}^1 dx~ W_1^{{\it v}_q^2}(x,Q^2,m^2)\,,
\\[2ex]
\label{eqn:2.23}
{\cal T}_2&=&-4 \int_{\sr}^1 dx~ W_1^{{\it a}_q^2}(x,Q^2,m^2)\,,
\\[2ex]
\label{eqn:2.24}
{\cal T}_3&=& \int_{\sr}^1 dx~\Bigg({\alpha_x^2 \over 2}
W_2^{{\it v}_q^2}(x,Q^2,m^2) -2 W_1^{{\it v}_q^2}(x,Q^2,m^2) \Bigg)\,,
\\[2ex]
\label{eqn:2.25}
{\cal T}_4&=& \int_{\sr}^1 dx ~\Bigg({\alpha_x^2 \over 2}
W_2^{{\it a}_q^2}(x,Q^2,m^2) -2 W_1^{{\it a}_q^2}(x,Q^2,m^2) \Bigg)\,,
\\[2ex]
\label{eqn:2.26}
{\cal T}_5&=& 2~\int_{\sr}^1 dx~\alpha_x W_6^{\{{\it v}_q,{\it a}_q\}}
(x,Q^2,m^2)\,,
\\[2ex]
\label{eqn:2.27}
{\cal T}_6&=& 2 \int_{\sr}^1 dx~\alpha_x W_6^{[{\it v}_q,{\it a}_q]}
(x,Q^2,m^2)\,,
\end{eqnarray}
The results for the longitudinal polarized structure functions are given by
\begin{eqnarray}
\label{eqn:2.28}
{\cal T}_{7,L_1}&=&\int_{\sr}^1 dx~\Big(
 -{\alpha_x \over 2}  W_3^{\{{\it v}_q,{\it a}_q\}}(x,Q^2,m^2) \Big)\,,
\\[2ex]
\label{eqn:2.29}
{\cal T}_{7,L_2} &=&  \int_{\sr}^1 dx~\Big( -{\alpha_x \over 2}
W_3^{\{{\it v}_q,{\it a}_q\}}(x,Q^2,m^2) +{\alpha_x^3 \over 8}
W_4^{\{{\it v}_q,{\it a}_q\}}(x,Q^2,m^2)
\nonumber\\[2ex]
&& +{\alpha_x \over 2} x W_5^{\{{\it v}_q,{\it a}_q\}}(x,Q^2,m^2) \Big)\,,
\\[2ex]
\label{eqn:2.30}
{\cal T}_{8,L}&=&\int_{\sr }^1 dx~\Big(
 -{\alpha_x \over 2}  W_3^{[{\it v}_q,{\it a}_q]}(x,Q^2,m^2) \Big)\,,
\\[2ex]
\label{eqn:2.31}
{\cal T}_{10,L} &=&  \int_{\sr}^1 dx~\Big( {\alpha_x^2 \over 4}
W_7^{{\it v}_q^2}(x,Q^2,m^2) -{\rho \over 4} W_8^{{\it v}_q^2}(x,Q^2,m^2)
\nonumber\\[2ex]
&& -{x \over 2} W_9^{{\it v}_q^2}(x,Q^2,m^2) \Big)\,,
\\[2ex]
\label{eqn:2.32}
{\cal T}_{11,L} &=& \int_{\sr}^1 dx~\Big( {\alpha_x^2 \over 4}
W_7^{{\it a}_q^2}(x,Q^2,m^2) -{\rho \over 4} W_8^{{\it a}_q^2}(x,Q^2,m^2)
\nonumber \\[2ex]
&& -{x \over 2} W_9^{{\it a}_q^2}(x,Q^2,m^2) \Big)\,.
\end{eqnarray}
Similar expressions are found for the transverse polarized structure functions
\begin{eqnarray}
\label{eqn:2.33}
{\cal T}_{7,T}&=& \int_{\sr}^1 dx \Big( -{\sr \over 2}  \alpha_x
W_5^{\{{\it v}_q, {\it a}_q\}}(x,Q^2,m^2) \Big)\,,
\\[2ex]
\label{eqn:2.34}
{\cal T}_{8,T}&=& \int_{\sr}^1 dx \Big( -{\sr \over 2}  \alpha_x
W_5^{[{\it v}_q, {\it a}_q]}(x,Q^2,m^2) \Big)\,,
\\[2ex]
\label{eqn:2.35}
{\cal T}_{10,T}&=& \int_{\sr}^1 dx~\Big( -{\sr \over 4} x
W_8^{{\it v}_q^2}(x,Q^2,m^2) -{\sr \over 2} W_9^{{\it v}_q^2}(x,Q^2,m^2) 
\Big)\,,
\\[2ex]
\label{eqn:2.36}
{\cal T}_{11,T}&=& \int_{\sr}^1 dx~\Big( -{\sr \over 4} x
W_8^{{\it a}_q^2}(x,Q^2,m^2) -{\sr \over 2} W_9^{{\it a}_q^2}(x,Q^2,m^2) \Big)
\,,
\end{eqnarray}
and the results for the normal polarized structure functions can be expressed
into the form
\begin{eqnarray}
\label{eqn:2.37}
{\cal T}_{9,N}&=& \int_{\sr}^1 dx \Big( {i \sr \over 8}  \alpha_x^2
W_{11}^{{\it v}_q^2}( x,Q^2,m^2) \Big)\,,
\\[2ex]
\label{eqn:2.38}
{\cal T}_{12,N}&=& \int_{\sr}^1 dx \Big( {i \sr \over 2}  \alpha_x
W_{10}^{\{{\it v}_q,{\it a}_q\}}( x,Q^2,m^2) \Big)\,,
\\[2ex]
\label{eqn:2.39}
{\cal T}_{13,N}&=& \int_{\sr}^1 dx \Big( {i \sr \over 2}  \alpha_x
W_{10}^{[{\it v}_q,{\it a}_q]}( x,Q^2,m^2) \Big)\,,
\end{eqnarray}
with
\begin{eqnarray}
\label{eqn:2.40}
\alpha_x=\sqrt{x^2-\rho}\,.
\end{eqnarray}
Analogous to deep inelastic lepton hadron scattering the leading contributions
to the unpolarized spin structure functions $W_i$ with $i=1,2,6$ in Eqs.
(\ref{eqn:2.22})-(\ref{eqn:2.27}) are of type twist two. 
The same holds for the polarized
structure function $W_3$. However $W_4$, $W_5$ and $W_7$, $W_8$, $W_9$,
$W_{10}$, $W_{11}$ contain
twist two as well as twist three contributions.
Notice that $W_8$ is due to the fact that the electroweak currents are
not conserved. Furthermore the twist three part cancels in the combinations
\begin{eqnarray}
\label{eqn:2.41}
xW_4+4\,W_5 \,,\qquad - xW_7+ 2\,W_9\,,
\end{eqnarray}
which means that the leading contributions in $m^2/Q^2$ to the longitudinal 
structure functions ${\cal T}_{i,L}$ in Eqs. (\ref{eqn:2.28})-(\ref{eqn:2.32}) 
are of twist two only. Notice that $W_8$ in the equations above is multiplied 
by $\rho=4m^2/Q^2$ so that this term vanishes in the limit $m \rightarrow 0$. 
The transverse parts ${\cal T}_{i,T}$ in Eqs. (\ref{eqn:2.33})-(\ref{eqn:2.36}) 
receive contributions from twist two as well as twist three. The same also
applies to the normal parts ${\cal T}_{i,N}$ in 
Eqs. (\ref{eqn:2.37})- (\ref{eqn:2.39}). A second feature of the above 
equations is that in the limit
$m \rightarrow 0$ all transverse and normal parts vanish whereas the 
longitudinal
parts ${\cal T}_{i,L}$ ($i=7,10,11$) and the unpolarized quantities
${\cal T}_i$ ($i=1-5$) tend to non zero values.
After having carried out the integration over $x$ in Eq. (\ref{eqn:2.12}) 
the differential cross section can be written as
\begin{eqnarray}
\label{eqn:2.42}
\frac{d~\sigma}{d~\Omega}(\lama,\lamb,W)= \frac{d~\sigma_U}{d~\Omega}
+ W_L \,\frac{d~\sigma_L}{d~\Omega}
+ \wnw \,\frac{d~\sigma_T}{d~\Omega} 
+ W_N \,\frac{d~\sigma_N}{d~\Omega}\,,
\end{eqnarray}
where $U$ represents the unpolarized
cross section with respect to the outgoing quark. The four cross sections on 
the right hand side of Eq. (\ref{eqn:2.42}) can be decomposed according to the
vector bosons which appear in the intermediate state i.e.
\begin{eqnarray}
\label{eqn:2.43}
{d~\sigma_k \over d \Omega}(\lama,\lamb) &=& N_c \alpha^2\left [
                  {d~\sigma_k^{(\gamma\gamma)} \over d \Omega}
                  +{d~\sigma_k^{(ZZ)} \over d \Omega}
                  +{d~\sigma_k^{(\gamma Z)} \over d \Omega} \right ]\,,
\,\, k=U,T,L,N\,.
\end{eqnarray}
The results for the photon-photon interference term can be written as
\footnote {Very often one replaces $\alpha$ by $G_FM_Z^2\sqrt 2 sw^2cw^2/\pi$
(see e.g. \cite{krz}, \cite{grkotu1}).} 
\begin{eqnarray}
\label{eqn:2.44}
{d\sigma_U^{(\gamma\gamma)} \over d \Omega} &=&
Q^2 \pgg {Q_e^2 Q_q^2 } \Bigg[ (1-\lama \lamb ) \Bigg( {1 \over 8} (1+\cos^2 \theta) 
{\cal T}_1 + {1 \over 4} \sin^2 \theta {\cal T}_3 \Bigg ) \Bigg]\,,
\\[2ex]
\label{eqn:2.45}
{d\sigma_L^{(\gamma\gamma)} \over d \Omega} &=&
Q^2 \pgg { Q_e^2 Q_q^2  } \Bigg[(\lama-\lamb )\cos \theta {\cal T}_{10,L}
\Bigg]\,,
\\[2ex]
\label{eqn:2.46}
{d\sigma_T^{(\gamma\gamma)} \over d \Omega} &=&
Q^2 \pgg {Q_e^2 Q_q^2 } \Bigg[ ( \lama-\lamb ){\cal T}_{10,T} \Bigg]\,,
\\[2ex]
\label{eqn:2.47}
{d\sigma_N^{(\gamma\gamma)} \over d \Omega} &=&
Q^2 \pgg {Q_e^2 Q_q^2 }\Bigg[ ( 1-\lama \lamb )\cos \theta \sin \theta 
{\cal T}_{9,N} \Bigg]\,.
\end{eqnarray}
The Z-Z interference term receives contributions from
\begin{eqnarray}
\label{eqn:2.48}
{d \sigma_U^{(ZZ)} \over d \Omega} &=&Q^2 { \pzz \over c_w^4 s_w^4 
} \Bigg[ \bigg\{ -2 \gve \gae 
(\lama-\lamb) + (\gve^2+\gae^2)(1-\lama \lamb)\bigg \}
\nonumber \\[2ex]
&& \times \bigg\{ {1 \over 8} (1+\cos^2 \theta)(\gvq^2  {\cal T}_1+\gaq^2 
{\cal T}_2) + {1 \over 4} \sin^2 \theta (\gvq^2 {\cal T}_3+\gaq^2 
{\cal T}_4) \bigg\}
\nonumber\\[2ex]
&& +\bigg\{ 2 \gve \gae (1-\lama \lamb)-(\gve^2+\gae^2)(\lama- \lamb)\bigg \}
\nonumber\\[2ex]
&& \times \bigg \{{1 \over 2} \gvq \gaq \, \cos \theta \, {\cal T}_5 \bigg\}
\Bigg ]\,,
\\[2ex]
\label{eqn:2.49}
{d \sigma_L^{(ZZ)} \over d \Omega} &=&Q^2 {\pzz \over c_w^4 s_w^4 }
\Bigg[ \bigg\{ 2 \gve \gae (\lama-\lamb)
- (\gve^2+\gae^2)(1-\lama \lamb)\bigg \}
\nonumber \\[2ex]
&&\times \bigg\{ \gvq \gaq
\bigg ( ( 1 + \cos^2 \theta){\cal T}_{7,L_1} + \sin^2 \theta {\cal T}_{7,L_2}
\bigg ) \bigg\}
\nonumber\\[2ex]
&& +\bigg\{ -2 \gve \gae (1-\lama \lamb)+(\gve^2+\gae^2) (\lama- \lamb)\bigg \}
\nonumber\\[2ex]
&& \times \bigg \{ \cos \theta \, \bigg ( \gvq^2 {\cal T}_{10,L}
+\gaq^2 {\cal T}_{11,L} \bigg ) \bigg\} \Bigg]\,,
\\[2ex]
\label{eqn:2.50}
{d \sigma_T^{(ZZ)} \over d \Omega} &=&Q^2 {\pzz \over c_w^4 s_w^4}
\Bigg[ \bigg\{ 2 \gve \gae (\lama-\lamb) - (\gve^2+\gae^2)(1-\lama \lamb)
\bigg \} 
\nonumber\\[2ex]
&& \times \bigg\{ \gvq \gaq \, \cos \theta \, {\cal T}_{7,T} \bigg\}
\nonumber\\[2ex]
&&+ \bigg\{ -2 \gve \gae (1-\lama \lamb) +(\gve^2+\gae^2) (\lama- \lamb)\bigg \}
\nonumber\\[2ex]
&&\times \bigg \{\gvq^2 {\cal T}_{10,T}+\gaq^2 {\cal T}_{11,T}
\bigg\} \Bigg]\,,
\\[2ex]
\label{eqn:2.51}
{d \sigma_N^{(ZZ)} \over d \Omega} &=&Q^2 { \pzz \over c_w^4 s_w^4 }
\Bigg[ \bigg\{- 2 \gve \gae (\lama-\lamb)
+ (\gve^2+\gae^2)(1-\lama \lamb)\bigg \}  
\nonumber\\[2ex]
&& \times \bigg\{ \gvq^2 \, \cos \theta \sin \theta \, {\cal T}_{9,N} \bigg \} 
\nonumber\\[2ex]
&& + \bigg\{ 2 \gve \gae (1-\lama \lamb)
-(\gve^2+\gae^2) (\lama- \lamb)\bigg \}
\nonumber\\[2ex]
&& \times \bigg\{ \gvq \gaq \, \sin \theta \, {\cal T}_{12,N} \bigg \} \Bigg]\,.
\end{eqnarray}
The photon-Z interference term consists of the following parts
\begin{eqnarray}
\label{eqn:2.52}
{d \sigma_U^{(\gamma Z)} \over d \Omega} &=&2 Q^2 {\Rgz \over c_w^2 s_w^2 } 
Q_e Q_q \Bigg[ \bigg\{  \gve (1-\lama \lamb) - \gae (\lama- \lamb)\bigg \} 
\nonumber\\[2ex]
&& \times \bigg\{ {1 \over 8} (1+\cos^2 \theta)\gvq  {\cal T}_1
           + {1 \over 4} \sin^2 \theta \gvq {\cal T}_3 \bigg\}
\nonumber\\[2ex]
&&  +\bigg\{- \gve (\lama- \lamb)+ \gae(1-\lama \lamb)\bigg \}
{1 \over 4} \gaq \, \cos \theta \, {\cal T}_5 \Bigg]
\nonumber\\[2ex]
&&+2 Q^2 { \Igz \over c_w^2 s_w^2} Q_e Q_q
\Bigg[ \bigg\{\gve (\lama-\lamb) -\gae (1-\lama \lamb) \bigg\}
\nonumber\\[2ex]
&& \times {1 \over 4} \gaq \, \cos \theta \, {\rm Im} \, {\cal T}_6 \Bigg ]\,,
\\[2ex]
\label{eqn:2.53}
{d \sigma_L^{(\gamma Z)} \over d \Omega}
&=&2 Q^2 {\Rgz \over c_w^2 s_w^2 } Q_e Q_q
\Bigg[ \bigg\{- \gve (1-\lama \lamb) + \gae (\lama- \lamb)\bigg \}
\nonumber\\[2ex]
&& \times \bigg\{ {1 \over 2} \gaq \bigg (
(1 + \cos^2 \theta ) {\cal T}_{7,L_1} + \sin^2 \theta {\cal T}_{7,L_2} \bigg )
\bigg \}
\nonumber\\[2ex]
&&     +\bigg\{  \gve (\lama- \lamb)- \gae(1-\lama \lamb)\bigg \}
            \gvq \,\cos \theta \,{\cal T}_{10,L}  \Bigg]
\nonumber\\[2ex]
&&+2 Q^2 { \Igz \over c_w^2 s_w^2 } Q_e Q_q
\Bigg[ \bigg\{\gve (1-\lama\lamb) -\gae (\lama-\lamb) \bigg\}
\nonumber\\[2ex]
&& \times \bigg\{ {1 \over 2} \gaq 
(1 + \cos^2 \theta ) {\rm Im} \,{\cal T}_{8,L}  \bigg \} \Bigg]\,,
\\[2ex]
\label{eqn:2.54}
{d \sigma_T^{(\gamma Z)} \over d \Omega}
&=&2 Q^2 {\Rgz \over c_w^2 s_w^2 } Q_e Q_q
\Bigg[ \bigg\{- \gve (1-\lama \lamb) + \gae (\lama- \lamb)\bigg \} 
\nonumber\\[2ex]
&& \times \bigg\{ {1 \over 2} \gaq \, \cos \theta \, {\cal T}_{7,T} \bigg \}
\nonumber\\[2ex]
&&     +\bigg\{  \gve (\lama- \lamb)- \gae(1-\lama \lamb)\bigg \}
            \gvq {\cal T}_{10,T}\bigg\} \Bigg]
\nonumber\\[2ex]
&&+2 Q^2 { \Igz \over c_w^2 s_w^2 } Q_e Q_q
\Bigg[ \bigg\{\gve (1-\lama\lamb)
-\gae (\lama-\lamb) \bigg\}
\nonumber\\[2ex]
&& \times  \bigg\{ {1 \over 2} \gaq \, \cos \theta\,{\rm Im} \,{\cal T}_{8,T}
\bigg \} \Bigg]\,,
\\[2ex]
\label{eqn:2.55}
{d \sigma_N^{(\gamma Z)} \over d \Omega}
&=&2 Q^2 {\Rgz \over c_w^2 s_w^2 } Q_e Q_q
\Bigg[ \bigg\{ \gve (1-\lama \lamb)
- \gae (\lama- \lamb)\bigg \}  
\nonumber\\[2ex]
&& \times \bigg\{\gvq \, \cos \theta \sin \theta \,{\cal T}_{9,N} \bigg \}
\nonumber\\[2ex]
&&     +\bigg\{- \gve (\lama- \lamb)+ \gae(1-\lama \lamb)\bigg \}
           {1 \over 2}  \gaq \,  \sin \theta \, {\cal T}_{12,N}  \Bigg]
\nonumber\\[2ex]
&&+2 Q^2 { \Igz \over c_w^2 s_w^2 } Q_e Q_q
\Bigg[ \bigg\{\gve (\lama-\lamb) -\gae (1-\lama\lamb) \bigg\}
\nonumber\\[2ex]
&& \times {1 \over 2} \gaq \,\sin \theta\, {\rm Im} \,{\cal T}_{13,N} \Bigg]\,.
\end{eqnarray}
The Born contributions (zeroth order in $\alpha_s$) to the unpolarized
quantities denoted by ${\cal T}_i^{(0)}$ ($i=1,2,3,5$) can be found in 
\cite{krz}, \cite{grkotu1}, \cite{rane1}.
The longitudinal ${\cal T}_{7,L_1}^{(0)}$, ${\cal T}_{i,L}^{(0)}$ 
($i=10,11$) and the transverse polarized structure functions
${\cal T}_{7,T}^{(0)}$ and ${\cal T}_{10,T}^{(0)}$ were calculated in 
\cite{krz}. In the last reference one also finds the Born contribution
to the normal polarized quantity given by ${\rm Im} {\cal T}_{13,N}^{(0)}$. 
Notice that the quantities ${\cal T}_i$ not mentioned above all vanish in 
the Born approximation. The first order QCD contributions to ${\cal T}_{9,N}$
and ${\cal T}_{12,N}$ are also computed in \cite{krz} but the corrections to 
the other structure functions ${\cal T}_i$ were neglected. The latter are
computed in \cite{kopitu}, \cite{grkotu1}, \cite{grkotu2} (longitudinal) and 
\cite{grko} (transverse and normal). The exceptions are the order $\alpha_s$
contributions to ${\rm Im} {\cal T}_6$ and ${\rm Im} {\cal T}_{8,L}$ 
which will be presented in this paper for the first
time. The second order QCD corrections are not known yet but for light quarks
like $u,d,s$ they can be computed for the longitudinal polarized structure
functions ${\cal T}_{i,L}$ ($i=7,10,11$) and the results are shown in the 
next section. Notice that the transverse and normal polarized structure 
functions ${\cal T}_{i,T}$ and ${\cal T}_{i,N}$ vanish for massless quarks.


\mysection{Computation of the corrections up to order $\alpha_s^2$.}
The partonic structure tensor defined in Eq. (\ref{eqn:2.13}) is
represented by the following perturbation series
\begin{eqnarray}
\label{eqn:3.1}
W_{\mu \nu}^{(V_1V_2)}&=& \sum_{k=0}^{\infty} \left ( {\alpha_s(\mu^2) 
\over 4 \pi}\right )^k  W_{\mu \nu}^{(V_1V_2),(k)}\,,
\end{eqnarray}
where $\mu$ denotes the renormalization scale.
The same expression for the perturbation series holds for the the structure
functions $W_i^{(V1V2)}$ and the functions ${\cal T}_i$ in Eqs. 
(\ref{eqn:2.21})-(\ref{eqn:2.39}). The structure tensor is determined
by the following reaction
\begin{eqnarray}
\label{eqn:3.2}
V(q) \rightarrow H(p) + \bar H (p_1) + l(p_2) \cdots l(p_n)\,,
\end{eqnarray}
where $l(p_i)$ ($i=2,3,\cdots n$) represent the momenta of the light
partons ((anti-)quarks and gluons) in the final state. If the matrix element
of the process above is given by  $M^{\mu\nu,(V1V2)}$
then the partonic tensor is obtained by integration over the multi-partonic
phase space which also includes the heavy anti-quark i.e.
\begin{eqnarray}
\label{eqn:3.3}
W_{\mu\nu}^{(V1V2)}= \prod_{i=1}^n \int {d^3 p_i \over (2 \pi)^3 2 p_i^0}
(2\pi)^4 \delta^{(4)} \left (q - p- \sum_{j=1}^n p_j \right ) 
M_{\mu\nu}^{(V1V2)} \,.
\end{eqnarray}
The computation of the phase space integrals is carried out in the rest
frame of the vector boson $V$ and proceeds in the way as is given in
\cite{rine1} where the integrals are performed up to order $\alpha_s^2$ in the 
case of massless quarks. It can be easily extended for massive quarks up
to order $\alpha_s$. However in second order the integrals for massive quarks
become very tedious and we are only able to present the results for
massless quarks.

In the Born approximation the zeroth order structure functions
$W_i^{(V1V2),(0)}$ are determined by the following process
\begin{eqnarray}
\label{eqn:3.4}
V(q) \rightarrow H(p) + \bar H (p_1)\,.
\end{eqnarray}
In this case the phase space integral is trivial and the partonic
structure functions,
presented in Eq. (\ref{eqn:A.1}), are proportional to $\delta(1-x)$. From
Eqs. (\ref{eqn:2.22})-(\ref{eqn:2.39}) we obtain for the unpolarized
structure functions
\begin{eqnarray}
\label{eqn:3.5}
{\cal T}^{(0)}_1&=& \beta~,  \qquad {\cal T}^{(0)}_2 = \beta^3 ~,
\qquad {\cal T}^{(0)}_3 = {1 \over 2} \beta \rho ~,
\qquad {\cal T}^{(0)}_4 =0 ~,\qquad {\cal T}^{(0)}_5 = \beta^2 ~,
\nonumber\\[2ex]
{\rm Im} {\cal T}^{(0)}_6&=&0~, \qquad \beta =\sqrt{1-\rho}\,.
\end{eqnarray}
The longitudinal polarized quantities in the Born approximation are given by
\begin{eqnarray}
\label{eqn:3.6}
{\cal T}^{(0)}_{7,L_1}&=& -{\beta^2 \over 4}  ~,
\qquad {\cal T}^{(0)}_{7,L_2}= 0 ~, 
\qquad {\rm Im}\,{\cal T}^{(0)}_{8,L}=0~,
\qquad {\cal T}^{(0)}_{10,L} = -{\beta \over 4 } ~,
\nonumber\\[2ex]
{\cal T}^{(0)}_{11,L}&=& -{1 \over 4} \beta^3  ~.
\end{eqnarray}
The transverse polarized quantities equal
\begin{eqnarray}
\label{eqn:3.7}
{\cal T}^{(0)}_{7,T}&=&-{\beta^2 \over 4} \sr ~,
\qquad {\rm Im}\,{\cal T}^{(0)}_{8,T}=0~,
\qquad {\cal T}^{(0)}_{10,T} = -{\beta \over 4 }\sr  ~,
\qquad {\cal T}^{(0)}_{11,T}= 0 ~.
\end{eqnarray}
For the normal polarized quantities we get 
\begin{eqnarray}
\label{eqn:3.8}
{\cal T}^{(0)}_{9,N}&=& 0~,
\qquad {\cal T}^{(0)}_{12,N}=0~,
\qquad {\rm Im}\,{\cal T}^{(0)}_{13,N}= {\beta^2 \over 4} \sr~,
\nonumber\\[2ex]
\beta &=&\sqrt{1-\rho}\,.
\end{eqnarray}
The order $\alpha_s$ corrections originate from the one-loop contributions
to the Born reaction in Eq. (\ref{eqn:3.4}) and the gluon bremsstrahlung 
process 
\begin{eqnarray}
\label{eqn:3.9}
V(q) \rightarrow H(p) + \bar H (p_1) + g(p_2)\,.
\end{eqnarray}
To facilitate the calculation we split the partonic tensor
into a virtual (VIRT), a soft gluon (SOFT) and a hard (HARD) gluon part i.e.
\begin{eqnarray}
\label{eqn:3.10}
W_{\mu \nu}^{(V_1V_2),(1)}&= W_{\mu \nu}^{(V_1V_2),VIRT} + 
W_{\mu \nu}^{(V_1V_2),SOFT} + W_{\mu \nu}^{(V_1V_2),HARD} \,, 
\end{eqnarray}
Starting with the virtual contribution the order $\alpha_s$ corrected
vector boson quark vertex reads
\begin{eqnarray}
\label{eqn:3.11}
\Gamma_{a,\mu}^{V,(1)}&=&i\Bigg[ \gamma_\mu (1+ {\cal C}_1)~{\it v}_a^V+
\gamma_5 \gamma_\mu (1+ {\cal C}_1 +2 {\cal C}_2)~{\it a}_a^V 
\nonumber\\[2ex]
&&+{(2 p_{\mu}-q_\mu)\over 2 m} {\cal C}_2~ {\it v}_a^V+
\gamma_5 {q_\mu\over 2 m} {\cal C}_3~ {\it a}_a^V \Bigg]\,, 
\qquad V=\gamma, Z\,.
\end{eqnarray}
Here the functions $C_i$ are given by
\begin{eqnarray}
\label{eqn:3.12}
{\rm Re}\,\,{\cal C}_1&=&{\alpha_s \over 4 \pi} C_F \Bigg[- \Bigg( {2-\rho 
\over \beta} \ln (t) +2 \Bigg) 
      \ln\left( {\lambda^2 \over m^2}\right) - 4 - 3 \beta \ln (t)
\nonumber\\[2ex]
&&+{2 -\rho \over \beta } \Bigg \{ - {1 \over 2} \ln^2 (t) +
      2 \ln (t) \ln(1-t) + 2  \Li (t) + {2 \pi^2 \over 3}  \Bigg \} \Bigg]\,,
\nonumber\\[2ex]
{\rm Re}\,\,{\cal C}_2&=&- {\alpha_s \over 4 \pi} C_F\Bigg[ {\rho \over \beta} 
     \ln (t) \Bigg] \,,
\nonumber\\[2ex]
{\rm Re}\,\,{\cal C}_3&=&{\alpha_s \over 4 \pi} C_F\Bigg[ (\rho - 3) 
{\rho \over \beta} \ln (t) -2\rho \Bigg]\,,
\nonumber\\[2ex]
{\rm Im}\,\, {\cal C}_1&=&{\alpha_s \over 4 \pi} \pi C_F 
\Bigg[ - \Bigg( {2-\rho \over \beta}\Bigg)  \ln\left( {\lambda^2 \over m^2}
\right) - 3 \beta +{2 -\rho \over \beta }\Bigg \{ - \ln (t) 
\nonumber\\[2ex]
&& + 2  \ln(1-t) \Bigg \} \Bigg]\,,
\nonumber\\[2ex]
{\rm Im}\,\, {\cal C}_2&=& {\alpha_s \over 4 \pi} \pi C_F
\Bigg[ {- \rho \over \beta} \Bigg]\,,
\nonumber\\[2ex]
{\rm Im}\,\, {\cal C}_3&=& {\alpha_s \over 4 \pi} \pi C_F\Bigg[ (3- \rho )
{\rho \over \beta} \Bigg]\,,
\nonumber\\[2ex]
t&=&\frac{1-\beta}{1+\beta}\,,
\end{eqnarray}
where $C_F$ denotes the colour factor which is equal to $C_F=(N_c^2-1)/2 N_c$.
In the equation above we have introduced a fictitious mass of the gluon 
$\lambda$ which is needed to regularize the infrared divergence. The 
ultraviolet divergences, which cancel in the expressions above, can be 
regularized with the cut-off method ($\int^{\Lambda^2}d^4k$) which is known 
from old textbooks on QED. In this way one avoids the intricacies of the
$\gamma_5$-matrix prescription \cite{ne2} which is characteristic of 
n-dimensional
regularization. The mass renormalization is carried out in the pole mass
scheme (on-shell renormalization). However one can also choose the
${\overline {MS}}$-scheme and carry out the analysis of the radiative 
corrections using the running mass (see e.g. \cite{grkotu2}).
The contributions from the Born reaction and the one-loop corrections
are given by
\begin{eqnarray}
\label{eqn:3.13}
W_{\mu \nu}^{(V_1V_2),(0)} + W_{\mu \nu}^{(V_1V_2),VIRT}
& = & {N_c \beta \over 32 \pi^2} Tr\Bigg[{(1+\gamma_5 \Slash s)\over 2}
(\Slash  p+m)\Gamma_{a,\mu}^{V_1,(1)} (\Slash  p_1-m)\tilde 
\Gamma_{a,\nu}^{V_2,(1)} \Bigg]
\nonumber\\[2ex]
&& \times   \delta(1-x)\,,
\end{eqnarray}
with $\tilde \Gamma_\mu = \gamma_0 \Gamma_\mu^\dagger \gamma_0$.
The soft gluon part of the partonic tensor is given by
\begin{eqnarray}
\label{eqn:3.14}
W_{\mu \nu}^{(V_1V_2),SOFT}&=&{N_c \beta \over 16 \pi^2} S^{SOFT} 
Tr\Bigg[ { (1+ \gamma_5 \Slash s)\over 2 }(\Slash  p+m)
\Gamma_{a,\mu}^{V_1,(0)} (\Slash  p_1-m)\Gamma_{a,\nu}^{V_2,(0)} \Bigg] 
\delta(1-x)\,,
\nonumber\\[1ex]
\end{eqnarray}
with
\begin{eqnarray}
\label{eqn:3.15}
S^{SOFT}&=&-{\alpha_s \over 8 \pi^2} C_F \int_0^\omega
          {d^3 p_2 \over E_2} \Bigg( {m^2 \over (p_1\cdot p_2)^2}
            +{m^2 \over (p\cdot p_2)^2}
            -{2 p \cdot p_1 \over (p_1\cdot p_2)(p\cdot p_2)}\Bigg)\,.
\end{eqnarray}
The above integral will be evaluated in the rest frame of the vector boson $V$
where $\omega$ is a cut off on the gluon energy $E_2$ which is taken to be
much smaller than the quark mass $m$ (i.e. $\omega \ll m$). This is the reason
why like in the case of the contribution from the Born reaction and the 
virtual corrections the soft gluon part is proportional to $\delta(1-x)$. 
The result is
\begin{eqnarray}
\label{eqn:3.16}
S^{SOFT}&=& {\alpha_s \over 4 \pi} C_F \Bigg[ 
      - 2 \ln\left({4 \omega^2 \over \lambda^2} \right) +2
      -{2-\rho \over \beta} \Bigg \{ \ln (t) - 2\Li(t) - 2\Li(-t)
\nonumber \\
&&      +\ln\left({4 \omega^2 \over \lambda^2} \right) \ln (t)
      - 2 \ln (t) \ln (1-t) - 2 \ln (t) \ln (1+t)
\nonumber \\
&&      + \ln^2 (t)+ {\pi^2 \over 6} \Bigg \} \Bigg]\,.
\end{eqnarray}
Addition of the virtual ($V$) and soft ($S$) gluon parts leads to the 
expressions $W_i^{(V_1V_2),V+S}$ in Eq. (\ref{eqn:A.3}).
Note that in the combination ${\cal C}^{V+S}={\cal C}_1+S^{SOFT}$
the gluon regulator mass $\lambda$ vanishes i.e.
\begin{eqnarray}
\label{eqn:3.17}
{\cal C}^{V+S}&=&{\alpha_s \over 4 \pi} C_F \Bigg[
       -\left(2+{2-\rho \over \beta} \ln(t) \right) 
             \ln\left({4 \omega^2 \over m^2}\right)
       -{2-\rho \over \beta} \Bigg \{-4 \Li(t)  
\nonumber \\ 
&&-2 \Li(-t)-4 \ln(t) \ln (1-t) -2 \ln(t) \ln(1+t)+{3 \over 2} \ln^2(t)
\nonumber \\ 
&&  -{\pi^2 \over2}  \Bigg \} -2 - {5-4 \rho \over \beta} \ln(t) \Bigg]\,.
\end{eqnarray}
The hard gluon part of the partonic tensor can be calculated in a 
straightforward way and the results are given in Eq. (\ref{eqn:A.4}).
Notice that the upper bound on the integral over
$x$ in Eq. (\ref{eqn:2.12}) for $W_{\mu \nu}^{(V_1V_2),HARD}$ is given 
by $1-2m\omega/Q^2$ so that the energy cut off $\omega$ is cancelled
between $W_i^{(V_1V_2),V+S}$ and $W_i^{(V_1V_2),HARD}$. 
The results for ${\cal T}_i$ are given in Eqs.
(\ref{eqn:B.1})-(\ref{eqn:B.18}). We tried to shorten these expression as much 
as possible by minimizing the number of independent polylogarithms. Note that 
there is a difference in sign
between ${\cal T}_{12,N}$ in Eq. (\ref{eqn:B.17}) and the equivalent expression
in Eq. (33) of \cite{grko}. However substitution of  ${\cal T}_{9,N}$ 
(Eq. (\ref{eqn:B.16})) and ${\cal T}_{12,N}$ (Eq. (\ref{eqn:B.17})) in the
cross section of Eq. (\ref{eqn:2.51}) leads to the same result as presented
in Eq. (16) in \cite{krz}. 

The computation of the order $\alpha_s^2$ corrections for massive quarks
is extremely tedious so that it has not been performed yet. The reason
is that the partonic structure functions for timelike processes like
$e^+e^-$-collisions cannot be written as the imaginary part of an
amplitude. This is in contrast to deep inelastic scattering where
$q^2$ is spacelike or the total cross section 
$\sigma_{tot}(e^+e^- \rightarrow `hadrons')$. In the latter case the cross 
section can be written as the imaginary part of the hadronic vacuum 
polarization function so that one is able to apply advanced methods to
compute these type of quantities (see e.g. \cite{chkust}). 
However for massless quarks we can compute the second order corrections
to the partonic structure functions contributing to the longitudinal
polarization. This is feasible using conventional
techniques as is shown in \cite{rine1}-\cite{rine2}. In the
case of massless quarks the computation is facilitated because one has the
following relations
\begin{eqnarray}
\label{eqn:3.18}
W_1^{{\it v}_q^2}=W_1^{{\it a}_q^2}&=&-{x \over 2}  W_3^{{\it v}_q{\it a}_q}\,,
\nonumber\\[2ex]
W_2^{{\it v}_q^2}=W_2^{{\it a}_q^2}&=&-{x \over 2}  W_4^{{\it v}_q{\it a}_q}
             -2 W_5^{{\it v}_q{\it a}_q}\,,
\nonumber\\[2ex]
W_6^{{\it v}_q{\it a}_q}&=&-{x \over 2}  W_7^{{\it v}_q^2,{\it a}_q^2}
             + W_9^{{\it v}_q^2,{\it a}_q^2}\,.
\end{eqnarray}
which follow from Eq. (\ref{eqn:2.13}) by putting $s=p/m$. Since $m=0$
the contributions coming from the vector currents ${\it v}_q^2$
lead to the same answer as those originating from the axial-vector
currents given by ${\it a}_q^2$. In this case the matrix element only
contains $\gamma$-matrices which anti-commute with the $\gamma_5$-matrix.
It also explains why the components proportional to ${\it v}_q^2$ and 
${\it a}_q^2$ which show up in the unpolarized structure functions are the
same as the contributions multiplying ${\it v}_q {\it a}_q$ appearing
in the polarized structure functions. The relations in Eq. (\ref{eqn:3.18})
break down for $m \not=0$ as will be discussed at the end of this section.
The structure functions $W_i$ in Eq. (\ref{eqn:2.13}) can
be related to the fragmentation functions $F_i^h(x,Q^2)$ (unpolarized see
\cite{rine1}, \cite{ne1}) and $g_i^h(x,Q^2)$ (polarized see \cite{rine2})
defined for
the process $e^+e^- \rightarrow h + "X"$. Here $h$ represents a hadron 
in the final state which originates from a light (anti-) quark.
Using the relations in Eqs. (\ref{eqn:2.22})- (\ref{eqn:2.32}), the 
quantities ${\cal T}_i$ can be expressed into the first moments of the 
non-singlet quark coefficient functions as follows
\begin{eqnarray}
\label{eqn:3.19}
{\cal T}_1&=&{\cal T}_2=\int_0^1dx~{\cal C}_{1,q}(x,\frac{Q^2}{\mu^2})\,,
\\[2ex]
\label{eqn:3.20}
{\cal T}_3&=&{\cal T}_4=\frac{1}{2} \int_0^1dx~\left (
{\cal C}_{1,q}(x,\frac{Q^2}{\mu^2})-{\cal C}_{2,q}(x,\frac{Q^2}{\mu^2})\right )
\,,
\\[2ex]
\label{eqn:3.21}
{\cal T}_5&=&\int_0^1dx~{\cal C}_{3,q}(x,\frac{Q^2}{\mu^2})\,,
\\[2ex]
\label{eqn:3.22}
{\cal T}_{7,L_1}&=&-\frac{1}{4} \int_0^1dx~\Delta {\cal C}_{5,q}
(x,\frac{Q^2}{\mu^2})\,,
\\[2ex]
\label{eqn:3.23}
{\cal T}_{7,L_2}&=&-\frac{1}{4} \int_0^1dx~\left (\Delta {\cal C}_{5,q}
(x,\frac{Q^2}{\mu^2})-\Delta {\cal C}_{4,q}(x,\frac{Q^2}{\mu^2}) \right )\,,
\\[2ex]
\label{eqn:3.24}
{\cal T}_{10,L}&=&{\cal T}_{11,L}=-\frac{1}{4} \int_0^1dx~\Delta {\cal C}_{1,q}
(x,\frac{Q^2}{\mu^2})\,,
\end{eqnarray}
Here $\mu$ represents the mass factorization scale as well as the 
renormalization scale. However since all coefficient functions above
are of the non-singlet type the dependence on the factorization scale 
vanishes while taking the first moments so that ${\cal T}_i$ only
depends on the renormalization scale. Because of the relations in  
Eq. (\ref{eqn:3.18}) the coefficient
functions above are not mutually independent. In \cite{rine2}, it was
demonstrated that one could derive 
the following relations between the polarized $\Delta {\cal C}_{i,q}$ 
($i=1,4,5$)
and the unpolarized ${\cal C}_{i,q}$ ($i=1,2,3$) coefficient functions.
They are given by
\begin{eqnarray}
\label{eqn:3.25}
\Delta {\cal C}_{1,q}={\cal C}_{3,q}\,, \quad \Delta {\cal C}_{4,q}=
{\cal C}_{2,q}\,, \quad \Delta {\cal C}_{5,q}= {\cal C}_{1,q}\,,
\end{eqnarray}
which means that in the case of massless quarks all quantities ${\cal T}_i$ are 
determined by three independent coefficient functions only.
The order $\alpha_s^2$ contributions originate from the processes
\begin{eqnarray}
\label{eqn:3.26}
&&V \rightarrow H + \bar H + g + g\,,
\nonumber\\[2ex]
&&V \rightarrow H + \bar H + H + \bar H\,,
\nonumber\\[2ex]
&&V \rightarrow H + \bar H + q + q\,,
\end{eqnarray}
which also includes the two-loop corrections to reaction (\ref{eqn:3.4}) and 
the one-loop corrections to process (\ref{eqn:3.9}).
The corresponding coefficient functions have been computed in 
\cite{rine1}-\cite{rine2} for massless quarks and their first moments are 
presented in \cite{rane1}. If we also include the contributions from
the lower order processes in (\ref{eqn:3.4}) and (\ref{eqn:3.9}) (see also
\cite{flsa}-\cite{stvo}) we obtain for massless quarks up to order $\alpha_s^2$
\begin{eqnarray}
\label{eqn:3.27}
{\cal T}_1^{m=0}\!\!\!\!&=&\!\!{\cal T}_2^{m=0}=1+ {\alpha_s(\mu^2) \over 4 \pi} 
   C_F \Big [1 \Big ] + \left({\alpha_s(\mu^2) \over 4 \pi} \right)^2 \Bigg[
             C_F^2 \left({ 7 \over 2}\right)
            +C_A C_F \left(-{11 \over 3} \ln\left({Q^2 \over \mu^2}\right)
            \right.
\nonumber \\[2ex]
&& \left.   +{347 \over 18} -44 \zeta(3) \right)
            +n_f C_F T_f \left({4 \over 3} \ln\left({Q^2 \over \mu^2}\right)
               -{62 \over 9} + 16 \zeta(3) \right) \Bigg]\,,
\\[2ex]
\label{eqn:3.28}
{\cal T}_3^{m=0}&=&{\cal T}_4^{m=0}\!\!\!=\!\!{\alpha_s(\mu^2) \over 4 \pi} 
C_F \Big [2 \Big ] + \left({\alpha_s(\mu^2) \over 4 \pi} \right)^2 \Bigg[
   C_F^2 \left(-5 \right) + C_A C_F \left(-{22 \over 3} \ln\left({Q^2 \over 
\mu^2}\right) \right.
\nonumber \\[2ex]
&& \left. +{380 \over 9} \right)  
+n_f C_F T_f \left({8 \over 3} \ln\left({Q^2 \over \mu^2}\right)
                  -{136 \over 9} \right)\Bigg ]\,,
\\[2ex]
\label{eqn:3.29}
{\cal T}_5^{m=0} &=&1+\left({\alpha_s(\mu^2) \over 4 \pi} \right)^2
\Bigg[C_A C_F \Big(-44 \zeta(3)\Big) +n_f C_F T_f \Big(16 \zeta(3)\Big) 
\Bigg]\,,
\\[2ex]
\label{eqn:3.30}
{\cal T}_{7,L_1}^{m=0}&=&-\frac{1}{4}{\cal T}_1^{m=0}\,,
\\[2ex]
\label{eqn:3.31}
{\cal T}_{7,L_2}^{m=0}&=&-\frac{1}{2}{\cal T}_3^{m=0}\,,
\\[2ex]
\label{eqn:3.32}
{\cal T}_{10,L}^{m=0}&=&{\cal T}_{11,L}^{m=0}=-\frac{1}{4}{\cal T}_5\,,
\end{eqnarray}
where we have chosen the ${\overline {MS}}$-scheme for the coupling constant
renormalization. In the above equations the colour factors are given by
$C_A=N_c$ and $T_f=1/2$ (for $C_F$ see below Eq. (\ref{eqn:3.12}). 
Furthermore $n_f$ is the number of light flavours and 
the scale $\mu$ represents the renormalization scale. Notice that
the sum ${\cal T}_1+{\cal T}_3$ is equal to $R_{e^+e^-}$ which is defined
by $\sigma_{tot}(e^+ +e^-\rightarrow hadrons)/\sigma_{tot}
(e^+ +e^-\rightarrow \mu^+ + \mu^-)$. Finally 
the results obtained for the spin dependent quantities ${\cal T}_{7,L_1}$, 
${\cal T}_{10,L}$, ${\cal T}_{11,L}$ depend on the way the limit 
$m \rightarrow 0$ is taken before or after the integration over the momenta.
This is revealed by a comparison of the order $\alpha_s$ corrections in
Eqs. (\ref{eqn:3.30})-(\ref{eqn:3.32}) with those obtained for the same
quantities in Eqs. (\ref{eqn:B.7}), (\ref{eqn:B.10}), (\ref{eqn:B.11}). 
We find the following relation
\begin{eqnarray}
\label{eqn:3.33}
{\cal T}_{7,L_1}^{(1),m \rightarrow 0}
+2 {\cal T}_{7,L_1}^{(0),m \rightarrow 0}= 
{\cal T}_{7,L_1}^{(1),m=0}\,, \quad
{\cal T}_{i,L}^{(1),m \rightarrow 0}+2 {\cal T}_{i,L}^{(0),m \rightarrow 0}= 
{\cal T}_{i,L}^{(1),m=0}\,, \quad i=10,11\,.
\nonumber\\
\end{eqnarray}
Therefore we expect that the same difference between the massless and massive
quark approach will happen in higher order. This difference originates from 
the property that the $\gamma_5$-matrix commutes with the mass term in the trace
(see e.g. Eq. (\ref{eqn:3.13})) contrary to the ordinary gamma-matrix with which
it anti-commutes. Hence the relations between the polarized and the
unpolarized partonic structure functions in Eq. (\ref{eqn:3.18}) will break 
down for the subleading terms. The origin of this phenomenon is explained
in \cite{grkotu2}
\footnote {A similar phenomenon has been observed in polarized deep inelastic 
lepton-hadron scattering for the structure function $\Delta g_1$ see
\cite{mene}.}
. It does not show up in the relations between the
${\it v}_q^2$- and ${\it a}_q^2$-parts of the unpolarized structure functions.
Therefore only the coefficient functions $\Delta {\cal C}_i$ ($i=1,4,5$)
in Eqs. (\ref{eqn:3.22})-(\ref{eqn:3.24}) will get an 
anomalous term while going from the massless to the massive quark approach. 
It also turns out that this term cancels in the combination 
$\Delta {\cal C}_5-\Delta {\cal C}_4$ so that ${\cal T}_{7,L_2}$ in Eq.
(\ref{eqn:3.23}) will be unaffected at least up to order $\alpha_s$. 
It is unlikely that the latter will also hold in higher order of perturbation 
theory. 

\mysection{Results}
In this section we will present the effect of the higher order QCD
corrections to the polarization of the quark in $e^+~e^-$-collisions. 
When both the incoming leptons as well as the outgoing quark are unpolarized 
the differential cross section is given by
\begin{eqnarray}
\label{eqn:4.1}
{d {\overline {\sigma}} \over d \Omega}\!\!&=&\!\!{1 \over 4}\sum_{\lama,\lamb}
      \Bigg({ d\sigma \over d \Omega}(\lama,\lamb,W)
       +{ d\sigma \over d \Omega}(\lama,\lamb,-W)\Bigg)
={1 \over 2}\sum_{\lama,\lamb} {d \sigma_U \over d \Omega}\,.
\end{eqnarray}
When the incoming leptons are unpolarized but the quark is polarized
the asymmetry is defined by
\begin{eqnarray}
\label{eqn:4.2}
{d \sigma_W \over d \Omega}\!\!&=&\!\!{1 \over 4}\sum_{\lama,\lamb}
      \Bigg({ d\sigma \over d \Omega}(\lama,\lamb,W)
       -{ d\sigma \over d \Omega}(\lama,\lamb,-W)\Bigg)
\nonumber\\[2ex]
&=&{1 \over 2}\sum_{\lama,\lamb}\Bigg ( W_L\,
{d \sigma_L \over d \Omega} + W_T\, {d \sigma_T \over d \Omega}
+  W_N\, {d \sigma_N \over d \Omega} \Bigg )\,.
\end{eqnarray}
In the case the electron is polarized the asymmetry becomes equal to
\begin{eqnarray}
\label{eqn:4.3}
{d \sigma_W(\lambda_{e^-})\over d \Omega}&=&{1 \over 2}
      \sum_{\lama} \Bigg({ d\sigma \over d \Omega}(\lama,\lamb,W)
       -{ d\sigma \over d \Omega}(\lama,\lamb,-W) \Bigg )
\nonumber\\[2ex]
&=&\sum_{\lama} \Bigg ( W_L\, {d \sigma_L \over d \Omega}
+ W_T\, {d \sigma_T \over d \Omega}
+  W_N\, {d \sigma_N \over d \Omega} \Bigg )\,.
\end{eqnarray}
When the positron is polarized and the electron is unpolarized we have the
relation
\begin{eqnarray}
\label{eqn:4.4}
{d \sigma_W(\lama)\over d \Omega}=
{d \sigma_W(-\lamb)\over d \Omega}
\end{eqnarray}
The polarizations of the quark are defined by
\begin{eqnarray}
\label{eqn:4.5}
&& P_k=  W_k {\sum_{\lama,\lamb} d\sigma_k/d\Omega 
\over \sum_{\lama,\lamb} d\sigma_U/d\Omega }\,, \qquad
P_k(\lamb)=\,W_k {\sum_{\lama} d\sigma_k/d\Omega \over 
\sum_{\lama} d\sigma_U/ d\Omega}\,, 
\nonumber\\[2ex]
&& k=L,T,N \,.
\end{eqnarray}
For the longitudinally polarized quark we choose the following spin vector
\begin{eqnarray}
\label{eqn:4.6}
\hat W = \hat n\,, \quad \rightarrow \quad  W_L=1\,,\quad W_T=0\,,\quad W_N=0\,.
\end{eqnarray}
For the transversely polarized quark the spin vector is chosen to be in the 
plane spanned by the electron and the quark momenta
\begin{eqnarray}
\label{eqn:4.7}
\hat W=\frac {\hat n \times (\hat n \times \hat q_{e^-})}{|\hat n \times
(\hat n \times \hat q_{e^-})|}\,, \quad \rightarrow \quad W_L=0\,, \quad
W_T=-\sin \theta\,, \quad \quad W_N=0\,.
\end{eqnarray}
For the quark polarisation which is directed normal to the plan spanned by the 
electron and the quark momenta we make the choice
\begin{eqnarray}
\label{eqn:4.8}
\hat W={\hat n \times \hat q_{e^-} \over |\hat n \times \hat q_{e^-}|}\,,
\quad \rightarrow \quad W_L=0\,,\quad W_T=0\,, \quad W_N=-1\,.
\end{eqnarray}
Notice that $\hat W$ is chosen in the same way as in \cite{grko}
but opposite the choice made in \cite{krz} \footnote{In the case the anti-quark
$\bar H$ is detected in the final state we have $P_k(\bar H)=-P_k(H)$ for 
$k=L,T$ but $P_N(\bar H)=P_N(H)$.}.
With the definitions above one can infer that on the Born level we
have the following properties. In the case of longitudinally polarized
quarks ($W_T=0, W_N=0$) one obtains for $W_L=\pm 1$
\begin{eqnarray}
\label{eqn:4.9}
&& \sum_{\lama}{ d\sigma \over d \Omega}(\lama,\lamb=W_L,W_L,\cos 
\theta=1) =\sum_{\lama} { d\sigma_U \over d \Omega}(\cos \theta=1) 
\nonumber\\[2ex]
&& \sum_{\lama}{ d\sigma \over d \Omega}(\lama,\lamb=-W_L,W_L,\cos 
\theta=1) =0
\nonumber\\[2ex]
&& \sum_{\lama}{ d\sigma \over d \Omega}(\lama,\lamb=-W_L,W_L,\cos 
\theta=-1) = \sum_{\lama} { d\sigma_U \over d \Omega}(\cos \theta=-1)
\nonumber\\[2ex]
&& \sum_{\lama}{ d\sigma \over d \Omega}(\lama,\lamb=W_L,,W_L,\cos 
\theta=-1) =0
\end{eqnarray}
The above relations hold for massive and massless quarks. From Eqs. 
(\ref{eqn:4.3}) and (\ref{eqn:4.5}) it follows
\begin{eqnarray}
\label{eqn:4.10}
P_L(\lamb=\pm 1,\cos \theta=\pm 1)=1 \qquad P_L(\lamb=\pm 1,\cos \theta=\mp 1)
=-1
\end{eqnarray}
For the transverse polarized cross section we also have a relation
which only holds at threshold $Q\rightarrow 2m$. It reads for $\theta= \pi/2$
($W_T=- 1$)
\begin{eqnarray}
\label{eqn:4.11}
&& \sum_{\lama}{ d\sigma \over d \Omega}(\lama,\lamb = W_T,W_T,\cos
\theta=0)\rightarrow \sum_{\lama} { d\sigma_U \over d \Omega}(\cos \theta=0)
\nonumber\\[2ex]
&& \sum_{\lama}{ d\sigma \over d \Omega}(\lama,\lamb=-W_T,W_T,\cos
\theta=0) \rightarrow 0
\end{eqnarray}
from which follows
\begin{eqnarray}
\label{eqn:4.12}
\mathop{\mbox{lim}}\limits_{\vphantom{
\frac{A}{A}} Q \rightarrow 2m}
 P_T(\lamb=\pm 1,\cos \theta=0)=\mp 1 
\end{eqnarray}
When $Q > 2m$ we get $|P_T|<1$. Far away from threshold the transverse
polarization becomes very small and tends to zero. All relations above
will be modified by QCD corrections as we will see below.

We will now discuss the effect of the higher order QCD corrections on the
polarizations of the quarks where we neglect any higher order effect coming 
from the
electro-weak sector. Our results are obtained by choosing the parameters
given in \cite{caso}. The electro-weak constants are: 
$M_Z=91.187~{\rm GeV/c^2}$, $\Gamma_Z=2.490~{\rm GeV/c^2}$ and 
$\sin^2 \theta_W=0.23116$. For the strong parameters we adopt 
$\Lambda_{\overline {MS}}=237~{\rm MeV/c}$ at $n_f=5$ so that the two-loop
corrected running coupling constant equals $\alpha_s(M_Z^2)=0.119$. Further
we take the renormalization scale $\mu=Q$. Furthermore we only study
up-quark, bottom- and top quark production for which the following masses
are chosen $m_u=0$, $m_b=4.5~{\rm GeV/c^2}$ and
$m_t=173.8~{\rm GeV/c^2}$. In our plots we will show the polarizations
computed in different orders of perturbation theory. Hence we follow the
notation in Eq. (\ref{eqn:3.1}) and define $P_k^{(i)}$ ($k=L,T,N$) to
be the order $\alpha_s^i$ contribution to the polarization.
Similarly we define $P_{k,i}$ to be the order $\alpha_s^i$ corrected
polarization. Finally we only show figures where the electron beam is 
polarized. In the case the positron is polarized and the electron beam is 
unpolarized the figures for $\lama=\pm 1$ are the same as those for 
$\lamb=\mp 1$.

In Fig. 1 we have plotted the Born (zeroth order $\alpha_s$) and the 
higher order corrections to the longitudinal polarization of the up-quark
at $Q=M_Z$. The computation is done in the massless quark approach were
the anomalous terms are absent.
Since the QCD corrections for unpolarized beams are very small we could
only show the order $\alpha_s^2$ corrected result $P_{L,2}$. For this case
we have plotted the ratios $P_{L,1}/P_{L,0}$ and $P_{L,2}/P_{L,1}$ in Fig. 2.
From the latter figure we infer that the order $\alpha_s$ corrections
do not exceed the 6 pro-mille level. The order $\alpha_s^2$ corrections
are at most 2 pro-mille. In the case of polarized beams they become larger. 
For $\cos \theta=\pm 1$ the order $\alpha_s$ and order $\alpha_s^2$ 
corrections are $8.5 \%$ and $2.5 \%$ respectively. 
Similar features are shown in Fig. 3 by bottom production at the same CM energy
which are computed in the massive quark approach
so that the anomalous terms are implicitly present. Here the corrections for 
polarized beams are larger than those obtained for the up-quark
and they amount to $25 \%$ at $\cos \theta=\pm 1$ but the corrections for 
unpolarized beams are so small that they could not be shown in the figure. 
In Fig. 4 we have studied the validity of the massless quark
approach ($m=0$) and the massive quark approach ($m \rightarrow 0$)
for $P_L^{(1)}$ in the case of bottom production at $Q=M_Z$ with unpolarized
lepton beams. To that order we have plotted the ratios of the various 
approaches with respect to the exact result computed for 
$m=m_b=4.5~{\rm GeV/c^2}$. As expected the massless approach given by 
$P_L^{(1)}(m=0)$ does not work very well (see the dotted curve in Fig. 4)
but also the massive approach $P_L^{(1)}(m\rightarrow 0)$ is rather bad
in particular near $\cos \theta =\pm 1$. It turns out that only for
$m_b<0.2~{\rm GeV/c^2}$ the difference between $P_L^{(1)}(m\rightarrow 0)$
and $P_L^{(1)}(m=0.2)$ is less than $9 \%$. For a comparison we have also
shown the ratio of the unpolarized cross sections 
$d^2\sigma^{(1)}(m\rightarrow 0)/d^2\sigma^{(1)}(m=4.5)$ where no anomalous
terms are present so that 
$d^2\sigma^{(1)}(m\rightarrow 0)=d^2\sigma^{(1)}(m=0)$. In this case the 
massless quark approach works rather well except near $\cos \theta =\pm 1$.
This is the justification for neglecting the bottom mass in the order
$\alpha_s^2$ contributions to the forward backward asymmetry and the shape
parameter in \cite{rane1}, \cite{case}. In Fig. 5 we also plotted the
longitudinal polarization of the top-quark which is produced at 
$Q=500~{\rm GeV}$. Like in the two previous cases the order $\alpha_s$ 
corrections to $P_L$ are extremely small even when the incoming 
leptons are polarized. Notice that for up-quark and bottom-quark production
at $Q=M_Z$ the $Z-Z$ interference term in Eq. (\ref{eqn:2.49}) dominates
the longitudinal polarization. However at $Q=500~{\rm GeV}$ it turns
out that the $\gamma-\gamma$ (Eq. (\ref{eqn:2.45})) and $\gamma-Z$ 
(Eq. (\ref{eqn:2.53})) interference terms become important too
although they partially cancel each other because the latter is negative.

For an analysis of the transverse $P_T$ and normal $P_N$ polarization we have 
to limit ourselves to heavy quark production since these quantities
vanish for massless quarks. In Fig. 6 we have shown $P_T$ at $Q=M_Z$
for bottom quarks. The QCD corrections are much larger than for $P_L$. This
also holds for unpolarized beams. On the other hand the absolute values for 
$P_T$ are rather small which is mainly due to the fact that the bottom
quark is produced far away from threshold $Q=M_Z \gg 2m_b$ so that 
$P_T\sim 2m_b/M_Z \ll 1$. For top-quark production at $Q=500~{\rm GeV}$ 
(see Fig. 7)
the situation is different. This quark is produced rather close to threshold
and the ratio $2m_t/Q\sim 0.7$ is large enough so that for polarized beams
the maximum value
of $|P_T|$ is close to one. However due to the difference $Q-2m_t$ there is a
shift from $\theta=\pi/2$ to larger angles (here about $\theta=2\pi/3$) 
where $P_T$ attains its maximum ($\lamb=-1$) or its minimum ($\lamb=1$).

The normal polarization $P_N$ is presented for the bottom quark ($Q=M_Z$)
in Fig. 8. From this figure we infer that $P_N$ is about a 
factor of five smaller than $P_T$ in Fig. 6.
Furthermore the difference between the Born
approximation $P_{N,0}$, which is wholly due to the quantity
${\cal T}_{13,N}^{(0)}$ in $d^2\sigma_N^{\gamma Z}$ (Eq. (\ref{eqn:2.55})), 
and the QCD corrected result $P_{N,1}$ is much larger than observed for 
$P_L$ and $P_T$ . Notice that the QCD corrections to
$P_N$ are not determined by the Z-peak in Eq. (\ref{eqn:2.51}) only but they 
also receive contributions from the $\gamma-Z$ interference term in 
Eq. (\ref{eqn:2.55}). Actually it is the structure function 
${\cal T}_{13,N}^{(1)}$ (see Eq. (\ref{eqn:B.18}) 
which is mainly responsible for the difference 
in Fig. 8 between $P_{N,0}$ and $P_{N,1}$ around $\cos \theta=0.8$ for
$\lamb=1$. When one is off-resonance the contributions in 
Eqs. (\ref{eqn:2.52})-(\ref{eqn:2.55})
which are proportional to ${\rm Im} P^{\gamma Z} \sim M_Z \Gamma_Z$ are
heavily suppressed. This is the reason why for top production at
$Q=500~{\rm GeV}$ (see Fig. 9) the Born approximation to the normal 
polarization given by $P_{N,0}$ is almost zero. Therefore in this case neither 
${\cal T}_{13,N}^{(0)}$ nor ${\cal T}_{13,N}^{(1)}$, both appearing in Eq. 
(\ref{eqn:2.55}), play any role anymore. Hence $P_N$ in Fig. 9, which is
smaller than $P_T$ in Fig. 7 by a factor of forty, is completely 
dominated by the QCD contributions coming from the structure functions 
${\cal T}_{9,N}^{(1)}$ (Eq. (\ref{eqn:B.16})) and ${\cal T}_{12,N}^{(1)}$ 
(Eq. (\ref{eqn:B.17})). 
This observation is important because besides of these corrections above,
$P_N$ can also receive contributions coming from CP-violating terms in the
neutral current sector which is a signal of new physics.
Hence it is important to compute the QCD corrections beyond order $\alpha_s$
which is not an easy task as we have discussed below Eq. (\ref{eqn:3.3}).
Finally we want to emphasize that for the top-quark all interference 
terms are equally important for the determination of $P_T$ and $P_N$. This
observation was already mentioned for $P_L$ above. This is because 
$t\bar t$-production occurs far above the Z-boson resonance so that the
dominance of the Z-propagator disappears.

As far as possible we have also made a comparison with some results obtained in
the literature. First we agree with the results presented for the top-quark for
polarized and unpolarized beams shown in Figs. 2, 9 in \cite{krz}. 
We also found agreement 
with the plots presented for $P_L$ and $P_T$ in \cite{grkotu1}- \cite{grkotu2} 
where the contributions ${\cal T}_6^{(1)}$ (Eq. (\ref{eqn:B.6}))
and ${\cal T}_{8,L}^{(1)}$ (Eq. (\ref{eqn:B.9})) were not taken into account.
However these contributions are negligible. On the other hand we disagree
with the normal polarization $P_N$ plotted for the bottom quark (Fig. 2b) and 
the top-quark (Fig. 3b) in \cite{grko}. This is due to the difference in
minus sign between ${\cal T}_{12,N}^{(1)}$ and the equivalent expression in
Eq. (33) of \cite{grko} which we have mentioned below Eq. (\ref{eqn:3.17}).
Note that the signs for ${\cal T}_{9,N}^{(1)}$ (Eq. (\ref{eqn:B.16})) 
and ${\cal T}_{12,N}^{(1)}$ (Eq. (\ref{eqn:B.17})) are the same as the
contributions appearing in Eq. (16) of \cite{krz}.

Before finishing this section we would like to comment on some 
results which are obtained in \cite{konapa} and \cite{pash}. Here one
has expressed $\hat W$ in Eq. (\ref{eqn:2.14}) into the following orthonormal 
basis (see also \cite{hage})
\begin{eqnarray}
\label{eqn:4.13}
\hat W=\sin \xi \cos \psi \hat n_1 
       +\sin \xi \sin \psi \hat n_2
       +\cos \xi \hat n_3\,,
\end{eqnarray}
with
\begin{eqnarray}
\label{eqn:4.14}
&& \hat n_1= {\hat n \times (\hat n \times e_3) \over
           |\hat n \times (\hat n \times e_3) |}\,,
\qquad 
\hat n_2= {\hat n \times  e_3 \over
           |\hat n \times e_3 |}\,,
\qquad
\hat n_3=-\hat n\,,
\nonumber\\[2ex]
&& e_1=(1,0,0)\,, \qquad e_2=(0,1,0)\,, \qquad e_3=(0,0,1)\,.
\end{eqnarray}  
Notice that $\hat n$ is the direction of the three-momentum of the observed
quark in the CM frame of the electron-positron pair whereas $\hat n_3=-\hat n$
points into the direction of the momentum $q=q_{e^+}+q_{e^-}$ in the
rest frame of the observed quark.
On the basis $e_i$ ($i=1,2,3$) the vectors $\hat n_i$ take the following form:
\begin{eqnarray}
\label{eqn:4.15}
\hat n_1&=&(\cos \theta \cos \phi, \cos \theta \sin \phi, -\sin \theta)\,,
\nonumber\\[2ex]
\hat n_2&=&(\sin \phi, -\cos \phi,0)\,,
\nonumber\\[2ex]
\hat n_3&=&-(\sin \theta \cos \phi, \sin \theta \sin \phi, \cos \theta)\,.
\end{eqnarray}
Substituting Eq. (\ref{eqn:4.13}) in Eq. (\ref{eqn:2.16}) and using the
parameterization for the quark momentum in (\ref{eqn:2.15}) 
the spin four-vector $s$ in the CM frame of the incoming leptons
has the following components
\begin{eqnarray}
\label{eqn:4.16}
s^0&=&-{Q \alpha_x \over 2 m}   \cos \xi \,, 
\nonumber\\[2ex]
s^1&=&-\left( {Q x \over 2 m} \right )
      \cos \xi \sin \theta \cos \phi 
      +\sin \xi \cos \psi \cos \theta \cos \phi
      +\sin \xi \sin \psi \sin \phi\,,
\nonumber\\[2ex]
s^2&=&-\left ( {Q x \over 2 m} \right )
      \cos \xi \sin \theta \sin \phi
      +\sin \xi \cos \psi \cos \theta \sin \phi
      -\sin \xi \sin \psi \cos \phi\,,
\nonumber\\[2ex]
s^3&=&-\left ( {Q x \over 2 m} \right )\cos \xi \cos \theta 
      - \sin \xi \cos \psi \sin \theta\,.
\end{eqnarray}
Similarly, using Eq.(\ref {eqn:4.13}) one finds that
\begin{eqnarray}
\label{eqn:4.17}
W_L&=&\nw =-\cos \xi\,,
\nonumber\\
\wnw&=&\hat W^3- \hat n . \hat W \cos \theta = -\sin \xi \cos \psi \sin \theta
\,,
\nonumber \\
W_N&=&\hat W^2 \cos \phi -\hat W^1 \sin \phi=-\sin \xi \sin \psi \,. 
\end{eqnarray}
Comparing the equation above with our choices of the polarization vector in 
Eqs. (\ref{eqn:4.6}) - (\ref{eqn:4.8}) we obtain
$\xi=0,\pi, \psi=arbitrary$ (longitudinal), $\xi=\pm \pi/2, \psi=0$ 
(transverse), and $\xi=\pm \pi/2, \psi=\pi/2$ (normal) respectively. 
Notice that the same angles were also adopted in \cite{kopitu}-\cite{grkotu2}.
A different choice was made made in \cite{konapa}, \cite{pash} where the values
of the angles are taken at $\psi=0$ and $\xi=arbitrary$. In this case only
the longitudinal and transverse polarization of the heavy quark can be studied 
since $W_N=0$. The reason for this choice is that the authors in 
\cite{konapa}, \cite{pash} did not include
the contributions coming from the imaginary parts of the virtual corrections
and the $Z$-boson propagator so that the normal polarization
vanishes. Furthermore in 
\cite{konapa} one has adopted three different bases for the spin vector
which are called the helicity, the beamline and the off-diagonal bases.  
The helicity bases is defined by the condition
\begin{eqnarray}
\label{eqn:4.18}
\cos \xi=\pm 1\,.
\end{eqnarray}
which is equivalent to our choice for the longitudinal polarization of
the quark. The beamline basis is given by
\begin{eqnarray}
\label{eqn:4.19}
\cos \xi =  {\cos \theta + \beta \over 1 + \beta \cos \theta}\,,
\qquad \beta=\sqrt{1-\rho}\,.
\end{eqnarray}
This condition only applies to the Born and the soft plus virtual
gluon contribution where there is no hard gluon emission. 
Here the spin of the quark is polarized along the positron 
momentum in the rest frame of the quark. The above equation is not
valid when there is hard gluon emission. In the later case one obtains
\begin{eqnarray}
\label{eqn:4.20}
\cos \xi =  {x \cos \theta + \alpha_x \over x + \alpha_x \cos \theta}\,,
\qquad \alpha_x=\sqrt{x^2-\rho}\,,
\end{eqnarray}
which reproduces the Eq.(\ref{eqn:4.19}) in the limit $x \rightarrow 1$
(soft gluon region) where $\alpha_x \rightarrow \beta$. 
Since $\cos \xi$ in Eq.(\ref{eqn:4.20}) is
a function of the integration variable $x$ in Eq. (\ref{eqn:2.12}) 
the choice above cannot be used in those
expressions where the $x$ integration is already done. 
Therefore in the case of hard gluon radiation the integral over $x$ is very 
complicated since this variable appears in the numerator as well as
denominator of the expression for $\cos \xi$. This will lead to a non-trivial 
dependence of the cross section on $\cos \theta$. Note that the choice
in Eq.(\ref{eqn:4.19}) at the level of hard gluon emission is at variance
with the interpretation that the spin of the top quark is polarized along 
the positron momentum direction in the rest frame of the quark.  
The third choice, called off-diagonal basis, corresponds to the case when
the like-spin configuration of the quark anti-quark pair vanishes identically.
As for the beamline basis this choice leads to an $x$-dependence of
$\cos \xi$ which is non-trivial so that it can be only applied to the Born and 
soft plus virtual gluon contribution.

Summarizing the results obtained in this paper we have presented the complete
first order QCD corrections to polarized (heavy) quark production. The most of 
these corrections were already calculated in \cite{kopitu}-\cite{grkotu2}.
We agree with these results except those obtained for the normal polarization
in \cite{grko}. Further we were able to compress the
expressions as far as possible by minimizing the number of polylogarithms.
Moreover we also computed the second order QCD corrections to the production
of light quarks
when the latter are longitudinally polarized. Furthermore we discovered
that the massless quark approach which was so successful to compute the
higher order corrections to unpolarized quantities in the kinematical regime
$Q\gg m$  failed in the case of the longitudinal polarization. As we have seen
for bottom-quark production this was not only due to the appearance of 
anomalous terms but also to the slow convergence to the zero mass limit.
Our results reveal that
the QCD corrections to the longitudinal polarization are rather small for
light as well as heavy quarks except for very small or very large values
of the scattering angle $\theta$. These corrections become more important
for the transverse polarization whereas in the case of the normal polarization
they completely dominate the process. For this reason it is very important
to compute the order $\alpha_s^2$ corrections to the normal polarization
which is a difficult task.\\[5mm]
\noindent
ACKNOWLEDGMENTS\\[3mm]
V. Ravindran would like to thank J. Bl\"umlein, H.S. Mani and S.D. Rindani
for discussions. The work of W.L. van Neerven was supported
by the EC network `QCD and Particle Structure' under contract
No.~FMRX--CT98--0194.


\appendix
\mysection{Appendix A}
Since the leptonic current is conserved for massless leptons, which implies 
$q_\mu {\cal L}^{\mu \nu}=q_\nu {\cal L}^{\mu \nu}=0$, we present only those 
partonic tensors which contribute to the cross section (see 
Eqs. (\ref{eqn:2.12}), (\ref{eqn:2.13})). 
In the Born approximation (\ref{eqn:3.4}) we obtain
\begin{eqnarray}
\label{eqn:A.1}
W_{1}^{(V_1V_2),(0)}&=&\Bigg(
           {\it v}_q^{V_1}~{\it v}_q^{V_2}
           \left [-{1 \over 4} \beta \right]
           +{\it a}_q^{V_1}~{\it a}_q^{V_2}
           \left [-{1 \over 4} \beta^3 \right] \Bigg) \delta(1-x)\,,
\nonumber\\[2ex]
W_{2}^{(V_1V_2),(0)}&=&-\Bigg(
           {\it v}_q^{V_1}~{\it v}_q^{V_2}
           +{\it a}_q^{V_1}~{\it a}_q^{V_2} \Bigg)
           \beta \delta(1-x)\,,
\nonumber\\[2ex]
W_{3}^{(V_1V_2),(0)}&=&{1 \over 2}\Bigg(
           {\it v}_q^{V_1}~{\it a}_q^{V_2}
           +{\it a}_q^{V_1}~{\it v}_q^{V_2} \Bigg)
           \beta \delta(1-x)\,,
\nonumber\\[2ex]
W_{4}^{(V_1V_2),(0)}&=&0\,,
\nonumber\\[2ex]
W_{5}^{(V_1V_2),(0)}&=&{1 \over 2}\Bigg(
           {\it v}_q^{V_1}~{\it a}_q^{V_2}
           +{\it a}_q^{V_1}~{\it v}_q^{V_2} \Bigg)
           \beta \delta(1-x)\,,
\nonumber\\[2ex]
W_{6}^{(V_1V_2),(0)}&=&{1 \over 2}\Bigg(
           {\it v}_q^{V_1}~{\it a}_q^{V_2}
           +{\it a}_q^{V_1}~{\it v}_q^{V_2} \Bigg)
           \beta \delta(1-x)\,,
\nonumber\\[2ex]
W_{7}^{(V_1V_2),(0)}&=&0\,,
\nonumber\\[2ex]
W_{8}^{(V_1V_2),(0)}&=&-\Bigg(
           {\it a}_q^{V_1}~{\it a}_q^{V_2} \Bigg) \beta \delta(1-x)\,,
\nonumber\\[2ex]
W_{9}^{(V_1V_2),(0)}&=&{1 \over 2}\Bigg(
           {\it v}_q^{V_1}~{\it v}_q^{V_2}
           +{\it a}_q^{V_1}~{\it a}_q^{V_2} \Bigg)\,,
           \beta \delta(1-x)
\nonumber\\[2ex]
W_{10}^{(V_1V_2),(0)}&=&{1 \over 2}\Bigg(
           {\it v}_q^{V_1}~{\it a}_q^{V_2}
           -{\it a}_q^{V_1}~{\it v}_q^{V_2} \Bigg)\,,
           \beta \delta(1-x)
\nonumber\\[2ex]
W_{11}^{(V_1V_2),(0)}&=&0\,,
\end{eqnarray}
with $V_1,V_2=\gamma,Z$.
For the order $\alpha_s$ correction (\ref{eqn:3.6}) it is convenient to 
decompose the partonic structure functions $W_i~(i=1-9)$ as follows 
\begin{eqnarray}
\label{eqn:A.2}
W_i^{(V_1V_2),(1)} &=& W_i^{(V_1V_2),V+S}+W_i^{(V_1V_2),HARD} \,,
\nonumber\\[2ex]
\!\!\!\!\mbox{with} \quad 
W_i^{(V_1V_2),V+S}&=&W_i^{(V_1V_2),VIRT}+ W_i^{(V_1V_2),SOFT}\,.
\end{eqnarray}
The soft plus virtual gluon contributions are given by
\begin{eqnarray}
\label{eqn:A.3}
W_1^{(V_1V_2),V+S} &=&-~{\beta \over 2}~ \Bigg[{\it v}_q^{V_1}~{\it v}_q^{V_2}
  ~ {\cal C}^{V+S} + {\it a}_q^{V_1}~{\it a}_q^{V_2}~ \beta^2 
  \Big (2~ {\rm Re}{\cal C}_2 + {\cal C}^{V+S} \Big) \Bigg] \delta(1-x)\,,
\nonumber\\[2ex]
W_2^{(V_1V2)),V+S} &=&-~2  \beta ~\Bigg [ {\it v}_q^{V_1}~{\it v}_q^{V_2}
\Big( {\rm Re}{\cal C}_2 + {\cal C}^{V+S} \Big) 
 +{\it a}_q^{V_1}~{\it a}_q^{V_2} 
\Big (2~{\rm Re}{\cal C}_2 + {\cal C}^{V+S} \Big ) \Bigg] \delta(1-x)\,,
\nonumber\\[2ex]
W_3^{(V_1V_2),V+S}&=& \beta~\Bigg [ 
  ({\it v}_q^{V_1}~{\it a}_q^{V_2}
  +{\it a}_q^{V_1}~{\it v}_q^{V_2})
  ~\Big( {\rm Re}{\cal C}_2 +{\cal C}^{V+S}\Big) 
\nonumber\\[2ex]
&&  -({\it v}_q^{V_1}~{\it a}_q^{V_2}
  -{\it a}_q^{V_1}~{\it v}_q^{V_2})
  ~i {\rm Im}{\cal C}_2 \Bigg] \delta(1-x)\,,
\nonumber\\[2ex]
W_4^{(V_1V_2),V+S}&=& 
   {2 \beta \over \rho}~\Bigg [ ({\it v}_q^{V_1}~{\it a}_q^{V_2}
  +{\it a}_q^{V_1}~{\it v}_q^{V_2}) ~{\rm Re}{\cal C}_2 
\nonumber\\[2ex]
&&  +  ({\it v}_q^{V_1}~{\it a}_q^{V_2} -{\it a}_q^{V_1}~{\it v}_q^{V_2})
   ~i {\rm Im}{\cal C}_2 \Bigg] \delta(1-x)\,,
\nonumber\\[2ex]
W_5^{(V_1V_2),V+S}&=& 
  {1 \over 2} \beta~\Bigg [ 
  ({\it v}_q^{V_1}~{\it a}_q^{V_2} +{\it a}_q^{V_1}~{\it v}_q^{V_2})
  ~\Big(-{1 \over \rho}~{\rm Re}{\cal C}_2 
  +3~{\rm Re}{\cal C}_2 +2~{\cal C}^{V+S}\Big) 
\nonumber\\[2ex]
&& + ({\it v}_q^{V_1}~{\it a}_q^{V_2} -{\it a}_q^{V_1}~{\it v}_q^{V_2})
  ~\Big(-{1 \over \rho}~i{\rm Im}{\cal C}_2 
  -~i {\rm Im}{\cal C}_2 \Big ) \Bigg] \delta(1-x)\,,
\nonumber\\[2ex]
W_6^{(V_1V_2),V+S}&=&  \beta \Bigg[ ({\it v}_q^{V_1}~{\it a}_q^{V_2}
  +{\it a}_q^{V_1}~{\it v}_q^{V_2}) ~\Big({\rm Re}{\cal C}_2 
+{\cal C}^{V+S}\Big) 
\nonumber\\[2ex]
&& - ({\it v}_q^{V_1}~{\it a}_q^{V_2} -{\it a}_q^{V_1}~{\it v}_q^{V_2})
  ~i {\rm Im}{\cal C}_2  \Bigg] \delta(1-x)\,,
\nonumber\\[2ex]
W_7^{(V_1V_2),V+S}&=& -~{\beta \over \rho} \Bigg[ {\it v}_q^{V_1}
    ~{\it v}_q^{V_2}~{\rm Re}{\cal C}_2 + {\it a}_q^{V_1}~{\it a}_q^{V_2}
    ~{\rm Re}{\cal C}_3 \Bigg] \delta(1-x)\,,
\nonumber\\[2ex]
W_8^{(V_1V_2),V+S}&=& -~\beta  \Bigg[ {\it a}_q^{V_1}~{\it a}_q^{V_2}
  ~\Big(4~ {\rm Re}{\cal C}_2 -{1 \over \rho}~ {\rm Re}{\cal C}_3 
+ 2~ {\cal C}^{V+S} \Big) \Bigg] \delta(1-x)\,,
\nonumber\\[2ex]
W_9^{(V_1V_2),V+S}&=& -~{\beta \over 2} \Bigg[ {\it v}_q^{V_1}~{\it v}_q^{V_2}
   ~\Big({1 \over \rho}~{\rm Re}{\cal C}_2 - {\rm Re}{\cal C}_2 
- 2~ {\cal C}^{V+S} \Big) 
\nonumber\\[2ex]
&&    + {\it a}_q^{V_1}~{\it a}_q^{V_2}~\Big(-4~ {\rm Re}{\cal C}_2 
+{1 \over \rho} ~ {\rm Re}{\cal C}_3 - 2~ {\cal C}^{V+S} \Big) \Bigg] 
\delta(1-x)\,,
\nonumber\\[2ex]
W_{10}^{(V_1V_2),V+S}&=& 
  {1 \over 2} \beta~\Bigg [ ({\it v}_q^{V_1}~{\it a}_q^{V_2}
  -{\it a}_q^{V_1}~{\it v}_q^{V_2})
  ~\Big(-{1 \over \rho}~ {\rm Re}{\cal C}_2 
  +3~{\rm Re}{\cal C}_2 +2~{\cal C}^{V+S}\Big) 
\nonumber\\[2ex]
&& + ({\it v}_q^{V_1}~{\it a}_q^{V_2} +{\it a}_q^{V_1}~{\it v}_q^{V_2})
  ~\Big(-{1 \over \rho}~i{\rm Im}{\cal C}_2 
  -~i {\rm Im}{\cal C}_2 ) \Bigg] \delta(1-x)\,,
\nonumber\\[2ex]
W_{11}^{(V_1V_2),V+S}&=& 
  -{2\beta \over \rho} ~\Bigg [ 
  {\it v}_q^{V_1}~{\it v}_q^{V_2}
  i{\rm Im}{\cal C}_2 \Bigg ] \delta(1-x)\,.
\end{eqnarray}
The hard gluon contributions are
\begin{eqnarray}
\label{eqn:A.4}
W^{(V_1V_2),HARD}_1\!\!&=&\!\!C_F {\it v}_q^{V_1}~{\it v}_q^{V_2}\Bigg[ 
 {1 \over \alpha_x \den^2 (1-x)} \Bigg( -20 \rho -11 \rho^2 -\rho^3   
\nonumber\\[2ex]
&&   + 8 x +18 \rho x  -2 \rho^2 x -2 \rho^3 x +8 x^2 +58 \rho x^2 
+23 \rho^2 x^2   
\nonumber\\[2ex]
&&  + \rho^3 x^2 -44 x^3 -74 \rho x^3 -8 \rho^2 x^3 +32 x^4 +18 
\rho x^4 -4 x^5 \Bigg)
\nonumber\\[2ex]
&&  +{ \ln(\xi) \over 4 \alpha_x^2 (1-x)} \Bigg(
     \rho -\rho^2 -2 \rho^2 x -2 x^2 +7 \rho x^2 +\rho^2 x^2
\nonumber\\[2ex]
&&     -2 \rho x^3 -2 x^4 \Bigg) \Bigg]       
\nonumber\\[2ex]
&&   +C_F {\it a}_q^{V_1}~{\it a}_q^{V_2} \Bigg[ {1 \over \alpha_x 
  \den^2 (1-x)} \Bigg( -20 \rho +25 \rho^2 +16 \rho^3   
\nonumber\\[2ex]
&& +2 \rho^4 +8 x   +10 \rho x -80 \rho^2 x -20 \rho^3 x +8 x^2 
+50 \rho x^2    
\nonumber\\[2ex]
&&  +55 \rho^2 x^2 -44 x^3 -34 \rho x^3  +2 \rho^2 x^3 +32 x^4 
  -6 \rho x^4 - 4 x^5 \Bigg)
\nonumber\\[2ex]
&&  +{ \ln(\xi) \over 4 \alpha_x^2 (1-x)} \Bigg(
    \rho +2 \rho^3 -8 \rho^2 x -2 x^2 + 7 \rho x^2 
+ 2 \rho x^3 -2 x^4 \Bigg) \Bigg]\,,
\nonumber\\[2ex]
W^{(V_1V_2),HARD}_2&=&C_F {\it v}_q^{V_1}~{\it v}_q^{V_2}\Bigg[ 
   {4 \over \alpha_x^3 \den^2 (1-x)} 
   \Bigg( 4 \rho +35 \rho^2 +17 \rho^3 +2 \rho^4   
\nonumber\\[2ex]
&&  +24 x -82 \rho x -110 \rho^2 x -22 \rho^3 x -40 x^2 
+186 \rho x^2 +85 \rho^2 x^2  
\nonumber\\[2ex]
&&  +\rho^3 x^2 +4 x^3 -126 \rho x^3 -8 \rho^2 x^3 +16 x^4 
+18 \rho x^4 -4 x^5 \Bigg)
\nonumber\\[2ex]
&&  -{ \ln(\xi) \over  \alpha_x^4 (1-x)} \Bigg(
    \rho -\rho^2 -2 \rho^3 +10 \rho^2 x +2 x^2 -13 \rho x^2 -\rho^2 x^2
\nonumber\\[2ex]
&&     +2 \rho x^3 +2 x^4 \Bigg) \Bigg] 
\nonumber\\[2ex]
&&+C_F {\it a}_q^{V_1}~{\it a}_q^{V_2}\Bigg[ {4 \over \alpha_x^3 \den^2 
(1-x)} \Bigg( 4 \rho +31 \rho^2 +18 \rho^3 +2 \rho^4  
\nonumber\\[2ex]
&& +24 x -106 \rho x  -108 \rho^2 x -24 \rho^3 x -40 x^2 +274 
\rho x^2  
\nonumber\\[2ex]
&&   +95 \rho^2 x^2  +2 \rho^3 x^2 +4 x^3 -242 \rho x^3 
-18 \rho^2 x^3 +16 x^4 +82 \rho x^4  
\nonumber\\[2ex]
&& +2 \rho^2 x^4 -4 x^5 -12 \rho x^5 \Bigg)
\nonumber\\[2ex]
&&   -{ \ln(\xi) \over  \alpha_x^4 (1-x)} \Bigg( \rho -2 \rho^2 -2 \rho^3 
+12 \rho^2 x +2 x^2-15 \rho x^2 -2 \rho^2 x^2 
\nonumber\\[2ex]
&&    +6 \rho x^3 +2 x^4 -2 \rho x^4 \Bigg) \Bigg] \,,
\nonumber\\[2ex]
W^{(V_1V_2),HARD}_3&=&
  {1\over 2}C_F ({\it v}_q^{V_1}~{\it a}_q^{V_2}
  +{\it a}_q^{V_1}~{\it v}_q^{V_2})
 \Bigg[ -{2 \over \alpha_x^3 \den^2 (1-x)} 
\Bigg( 68 \rho^2 +35 \rho^3 
\nonumber\\[2ex]
&&+4 \rho^4 -128 \rho x 
-208 \rho^2  x -45 \rho^3 x + 48 x^2 +332 \rho x^2 
\nonumber\\[2ex]
&&+172 \rho^2 x^2 +5 \rho^3 x^2  
-112 x^3  -304 \rho x^3 -52 \rho^2 x^3 -3 \rho^3 x^3 
\nonumber\\[2ex]
&&+104 x^4 +156 \rho x^4 
+24 \rho^2 x^4 -64 x^5 -56 \rho x^5 +24 x^6 \Bigg)
\nonumber\\[2ex]
&&   -{ \ln(\xi) \over  2 \alpha_x^4 (1-x)} \Bigg( 3 \rho^2 
+4 \rho^3 -2 \rho x -21 \rho^2 x +18 \rho x^2 +5 \rho^2 x^2  
\nonumber\\[2ex]      
&& -4 x^3 -2 \rho x^3 -3 \rho^2 x^3+6 \rho x^4 -4 x^5 \Bigg) \Bigg] \,,
\nonumber\\[2ex]
W^{(V_1V_2),HARD}_4&=& {1 \over 2}C_F ({\it v}_q^{V_1}~{\it a}_q^{V_2}
     +{\it a}_q^{V_1}~{\it v}_q^{V_2}) \Bigg[ {8 \over \alpha_x^7 \den^2 } 
\Bigg( 64 \rho^2 +124 \rho^3 
\nonumber\\[2ex]
&&+45 \rho^4 +4 \rho^5 
-528 \rho^2 x -404 \rho^3 x -63 \rho^4 x +112 \rho x^2 
\nonumber\\[2ex]
&&+976 \rho^2 x^2 +319 \rho^3 x^2 
+6 \rho^4 x^2 +96 \rho x^3 -432 \rho^2 x^3 -18 \rho^3 x^3 
\nonumber\\[2ex]
&&- 176 x^4 -764 \rho x^4 
-158 \rho^2 x^4 -9 \rho^3 x^4 +432 x^5 +756 \rho x^5 
\nonumber\\[2ex]
&&+75 \rho^2 x^5 -336 x^6   
-206 \rho x^6 -\rho^2 x^6 +80 x^7+6 \rho x^7 \Bigg) 
\nonumber\\[2ex]
&&     -{ 2 \ln(\xi) \over  \alpha_x^6 } \Bigg(
       13 \rho^2 +4 \rho^3 -18 \rho x -39 \rho^2 x + 58 \rho x^2
       +10 \rho^2 x^2  
\nonumber\\[2ex]
&&   -12 x^3 -21 \rho x^3 +4 x^4 +\rho x^4 \Bigg) \Bigg]\,,
\nonumber\\[2ex]
W^{(V_1V_2),HARD}_5&=&
   {1 \over 2}C_F ({\it v}_q^{V_1}~{\it a}_q^{V_2}
   +{\it a}_q^{V_1}~{\it v}_q^{V_2})
   \Bigg[ 
   -{2 \over \alpha_x^3 \den^2 (1-x)} 
  \Bigg( 80 \rho +96 \rho^2 
\nonumber\\[2ex]
&&+37 \rho^3  
+ 4 \rho^4 -96 x -440 \rho x - 272 \rho^2 x -46 \rho^3 x +400 x^2 
\nonumber\\[2ex]
&& +736 \rho x^2 +193 \rho^2 x^2 + \rho^3 x^2  -568 x^3 -446 \rho x^3 
-14 \rho^2 x^3   
\nonumber\\[2ex]
&& +320 x^4 +76 \rho x^4 +\rho^2 x^4 -56 x^5 -6 \rho x^5 \Bigg)
\nonumber\\[2ex]
&&     -{ \ln(\xi) \over  2 \alpha_x^4 (1-x)} \Bigg( 4 \rho + 5 \rho^2 
+4 \rho^3 -26 \rho x -22 \rho^2 x +8 x^2 
\nonumber\\[2ex]
&& +49 \rho x^2 +\rho^2 x^2 -16 x^3 -8 \rho x^3 
+ \rho x^4 \Bigg) \Bigg]\,,
\nonumber\\[2ex]
W^{(V_1V_2),HARD}_6&=&
    {1 \over 2}C_F ({\it v}_q^{V_1}~{\it a}_q^{V_2}
    +{\it a}_q^{V_1}~{\it v}_q^{V_2})
   \Bigg[ {8 \over \alpha_x \den^2 (1-x)} \Bigg( -8 +12 \rho 
\nonumber\\[2ex]
&&+8 \rho^2 +\rho^3  +12 x -31 \rho x -9 \rho^2 x -2 x^2 +18 \rho x^2 
+\rho x^3 -2 x^4 \Bigg)
\nonumber\\[2ex]
&&     +{2 \ln(\xi) \over  \alpha_x^2 (1-x)} \Bigg(
       \rho^2+x-3 \rho x +x^3 \Bigg) \Bigg]\,,
\nonumber\\[2ex]
W^{(V_1V_2),HARD}_7&=&C_F {\it v}_q^{V_1}~{\it v}_q^{V_2} 
    \Bigg [{2 \over \alpha_x^3 \den^2 } \Bigg(
       -64 \rho -36 \rho^2 -4 \rho^3 +96 x 
\nonumber\\[2ex]
&& +172 \rho x  +39 \rho^2 x -200 x^2 -102 \rho x^2 +\rho^2 x^2
       +104 x^3 -6 \rho x^3 \Bigg)
\nonumber\\[2ex]
&&     +{ \ln(\xi) \over 2 \alpha_x^4 } \Bigg(
        -4 \rho -4 \rho^2 +15 \rho x -8 x^2 +\rho x^2 \Bigg) \Bigg]
\nonumber\\[2ex]
&& +C_F {\it a}_q^{V_1}~{\it a}_q^{V_2} \Bigg[{2 \over \alpha_x^3 \den^2 } 
\Bigg( -64 \rho -16 \rho^2 -\rho^3 +96 x 
\nonumber\\[2ex]
&&  +132 \rho x  +\rho^2 x -3 \rho^3 x -200 x^2 -46 \rho x^2 
+19 \rho^2 x^2
\nonumber\\[2ex]
&&        +120 x^3 -22 \rho x^3 -16 x^4 \Bigg)
\nonumber\\[2ex]
&&     +{ \ln(\xi) \over 2 \alpha_x^4 } \Bigg( -4 \rho -\rho^2 +15 \rho x 
-3 \rho^2 x -8 x^2+\rho x^2 \Bigg) \Bigg]\,,
\nonumber\\[2ex]
W^{(V_1V_2),HARD}_8&=&C_F {\it a}_q^{V_1}~{\it a}_q^{V_2} 
     \Bigg [-{2 \over \alpha_x \den^2 (1-x)} 
 \Bigg( 52 \rho +31 \rho^2 +4 \rho^3 -56 x 
\nonumber\\[2ex]
&&  -130 \rho x -34 \rho^2 x +112 x^2 +72 \rho x^2 -\rho^2 x^2 
       -56 x^3 +6 \rho x^3 \Bigg) 
\nonumber\\[2ex]
&&   +{ \ln(\xi) \over 2 \alpha_x^2 (1-x)} \Bigg(
    \rho -4 \rho^2 +10 \rho x -8 x^2 +\rho x^2 \Bigg) \Bigg]\,,
\nonumber\\[2ex]
W^{(V_1V_2),HARD}_9&=&C_F {\it v}_q^{V_1}~{\it v}_q^{V_2} \Bigg [
    -{ 1 \over \alpha_x \den (1-x)} 
  \Bigg( -8 -17 \rho -4 \rho^2  
\nonumber\\[2ex]
&&   +22 x +18 \rho x  -4 x^2+3 \rho x^2 -10 x^3 \Bigg) \Bigg]
\nonumber\\[2ex]
&&   +{ \ln(\xi) \over 4 \alpha_x^2 (1-x)} \Bigg( \rho +4 \rho^2 -4 x 
     -10 \rho x +8 x^2 -3 \rho x^2+4 x^3 \Bigg) \Bigg]
\nonumber\\[2ex]
&& +C_F {\it a}_q^{V_1}~{\it a}_q^{V_2} \Bigg [{ 1 \over \alpha_x 
 \den^2 (1-x)} \Bigg( 32 -4 \rho + \rho^2 -120 x 
\nonumber\\[2ex]
&&  +62 \rho x  +30 \rho^2 x+4 \rho^3 x +120 x^2 -144 \rho x^2 
-35 \rho^2 x^2
\nonumber\\[2ex]
&&    -8 x^3 +86 \rho x^3 -24 x^4 \Bigg)
\nonumber\\[2ex]
&&     +{ \ln(\xi) \over 4 \alpha_x^2 (1-x)} \Bigg( \rho -4 x 
-2 \rho x +4 \rho^2 x +8 x^2 
\nonumber\\[2ex]
&& -11 \rho x^2 +4 x^3 \Bigg) \Bigg]\,,
\nonumber\\[2ex]
W^{(V_1V_2),HARD}_{10}&=& 
{1 \over 2}C_F ({\it v}_q^{V_1}~{\it a}_q^{V_2} 
               -{\it a}_q^{V_1}~{\it v}_q^{V_2}) \Bigg [
   {2 \over \alpha_x \den^2 (1-x)}
    \Bigg( 76 \rho +33 \rho^2 
\nonumber\\[2ex]
&&        +4 \rho^3 -72 x -190 \rho x
          -38 \rho^2 x + 144 x^2 +120 \rho x^2 +\rho^2 x^2 
\nonumber\\[2ex]
&&  -72 x^3 -6 \rho x^3\Bigg)  +{\ln(\xi) \over 2 \alpha_x^2 (1-x)}
     \Bigg(\rho +4 \rho^2 -14 \rho x +8 x^2 + \rho x^2\Bigg) \Bigg]\,,
\nonumber\\[2ex]
W^{(V_1V_2),HARD}_{11}&=& 0\,.
\end{eqnarray}
In the expressions above we have introduced the following notations
\begin{eqnarray}
\label{eqn:A.6}
\alpha_x=\sqrt{x^2-\rho}\,, \qquad
\xi = { \rho -2 x -2 \alpha_x \over  \rho -2 x +2 \alpha_x }\,.
\end{eqnarray}

\mysection{Appendix B}
In this appendix we present the order $\alpha_s$ contributions to the
structure functions ${\cal T}_i$ Eqs. (\ref{eqn:2.21})-(\ref{eqn:2.32}) 
computed in section 3. The unpolarized parts are given by
\begin{eqnarray}
\label{eqn:B.1}
{\cal T}^{(1)}_1&=& C_F\Bigg [{1 \over 2} \rho(1+\rho) F_1 + \sr (1-3 \rho) F_2
+2 (8-5 \rho -\rho^2) F_3 + 2 \beta F_4 
\nonumber\\[2ex]
&& +{1 \over 2} \beta (2+ 13 \rho)+{1 \over 2} (64 -39 \rho -7 \rho^2) \Li(t)
 +{1 \over 4} (-48+36 \rho -5 \rho^2) \ln(t)
\nonumber\\[2ex]
&& +2(4-3 \rho-\rho^2)  \ln(t) \ln(1+t) \Bigg ]\,,
\\[2ex]
\label{eqn:B.2}
{\cal T}^{(1)}_2&=& C_F \Bigg [{1 \over 2}\rho (1+ 2\rho)F_1 + \sr (1-4 \rho) 
F_2 +2 (8-13 \rho +2 \rho^2) F_3 + 2 \beta (1-\rho) F_4 
\nonumber\\[2ex]
&&  +{1 \over 2}\beta (2+\rho) +{1 \over 2} (64 -103 \rho +18 \rho^2) \Li(2)
   + {1 \over 4}(-48 +60 \rho -9 \rho^2) \ln(t)
\nonumber \\[2ex]
&&   +2 (4-7 \rho) \ln(t) \ln(1+t) \Bigg ]\,,
\\[2ex]
\label{eqn:B.3}
{\cal T}^{(1)}_3&=& C_F \Bigg [ -{1 \over 2} \rho (1+\rho) F_1 
+ \sr (-1+3 \rho) F_2 + 2 \rho (5-\rho) F_3 +\beta \rho F_4 +2 \beta (1-\rho)
\nonumber\\[2ex]
&&  +{1 \over 2} \rho (39 -9 \rho) \Li(t) +\rho (-7+3 \rho) \ln(t) 
+6 \rho \ln(t) \ln(1+t)\Bigg ]\,,
\\[2ex]
\label{eqn:B.4}
{\cal T}^{(1)}_4&=&C_F \Bigg [ -{1 \over 2} \rho (1+2 \rho) F_1 
+ \sr (-1+4 \rho) F_2 +2 \rho (1+2 \rho) F_3  
\nonumber\\[2ex]
&& +{1 \over 4} \beta (8 -38 \rho+3 \rho^2) +{7 \over 2} \rho (1+2 \rho) \Li(t)
 +{1 \over 8} \rho (-32 +8 \rho -3 \rho^2) \ln(t)
\nonumber\\[2ex]
&&   +2 \rho (1+2 \rho) \ln(t) \ln(1+t) \Bigg]\,,
\\[2ex]
\label{eqn:B.5}
{\cal T}^{(1)}_5&=&C_F \Bigg [ 4 \beta (-2+\rho) G_1 + 2 (4-5 \rho) G_2
   -4 \sr +4 \rho +8(-1+\rho) \ln(1+t)
\nonumber\\[2ex]
&&  + 8 \ln(1+t-\sqt) +16 (-1+\rho)  \ln(1-\sqt)
\nonumber\\[2ex]
&&    +2 (2-4 \rho +3 \beta  \rho -2 \beta) \ln(t) \Bigg]\,,
\\[2ex]
\label{eqn:B.6}
{\rm Im}{\cal T}_6^{(1)}&=& C_F \Bigg[ 2  \pi \rho \beta \Bigg]  \,. 
\end{eqnarray}
The longitudinal polarized structure functions are equal to
\begin{eqnarray}
\label{eqn:B.7}
{\cal T}_{7,L_1}^{(1)}&=& C_F \Bigg [\beta (2 -\rho) G_1 -{1 \over 4} 
(8+2 \rho +3 \rho^2) G_2 -{ \sr \over 8} (8 + 29 \rho)+ {1 \over 8}(2 +35 \rho)
\nonumber \\[2ex]
&& +2 (1-\rho) \ln(1+t) + {1 \over 16} (-32 +60 \rho -17 \rho^2) \ln(1+t-\sqt)
\nonumber \\[2ex]
&&+4 (1-\rho) \ln(1-\sqt)+{1 \over 32} (-32+4 \rho +32 \beta
   + 72 \rho \beta +17 \rho^2 )\ln(t)\Bigg]\,,
\nonumber\\
\\[2ex]
\label{eqn:B.8}
{\cal T}_{7,L_2}^{(1)}&=&  C_F \Bigg [ {\rho \over 2} (10 + 3\rho) G_2
  +{\sr \over 2}(8+13 \rho) - {1 \over 2} (2 +19 \rho)
\nonumber\\[2ex]
&&+{\rho \over 4} (-24 + 7 \rho) \ln(1+t-\sqt)
+ {\rho \over 8} (24 -52 \beta -7 \rho) \ln(t) \Bigg]\,,
\\[2ex]
\label{eqn:B.9}
{\rm Im}{\cal T}_{8,L}^{(1)}&=& C_F \Bigg [-{\pi \over 2} \rho \beta \Bigg]\,,
\\[2ex]
\label{eqn:B.10}
{\cal T}_{10,L}^{(1)}&=& C_F \Bigg [ {5 \rho \over 8}  F_1 -
{\sr \over 4} (4 +\rho) F_2 -{1 \over 2} (8+\rho) F_3 -{\beta \over 2} F_4 
+ {\beta \over 2}
\nonumber \\[2ex]
&&-{1 \over 8} (64 + 3 \rho) {\cal L}i_2(t) + {1 \over 4} (12 - 3 \rho) \ln(t)
-{1 \over 2} (4 + 3 \rho) \ln(t) \ln(1+t) \Bigg]\,,
\nonumber\\
\\[2ex]
\label{eqn:B.11}
{\cal T}_{11,L}^{(1)}&=& C_F \Bigg [{\rho \over 8} (5-\rho) F_1 -\sr F_2
      +{1 \over 2} (-8 + 7 \rho -3 \rho^2) F_3 + { \beta \over 2} (-1 +\rho)F_4
\nonumber  \\[2ex]
&&   +{\beta \over 2} (1+ 3 \rho)
+ {1 \over 8} (-64+61 \rho -25 \rho^2) {\cal L}i_2(t)
+{1 \over 4} (12-9 \rho + \rho^2) \ln(t)
\nonumber  \\[2ex]
&&+ {1 \over 2} (-4+\rho-\rho^2) \ln(t) \ln(1+t)\Bigg]\,.
\end{eqnarray}
The transverse polarized structure functions are
\begin{eqnarray}
\label{eqn:B.12}
{\cal T}_{7,T}^{(1)}&=&C_F \Bigg [ \sr \beta(2-\rho) G_1
- {\sr \over 4} (16+7 \rho) G_2 +{\sr \over 8} (48 +17 \rho)
-{\rho \over 8} (62 +3 \rho)
\nonumber \\[2ex]
&&  +2 \sr (1-\rho) \ln(1+t)+{\sr \over 16} (40+2 \rho-3 \rho^2) \ln(1+t-\sqt)
\nonumber \\[2ex]
&&  +4 \sr (1-\rho) \ln(1-\sqt) +{\sr \over 32} (-104 +62 \rho +168 \beta
+16 \rho \beta
\nonumber\\[2ex]
&& +3 \rho^2) \ln(t) \Bigg]\,.
\\[2ex]
\label{eqn:B.13}
{\rm Im}{\cal T}_{8,T}^{(1)}&=& C_F \Bigg [- {\pi \over 4}\sr \beta (1+\rho) 
\Bigg]\,,
\\[2ex]
\label{eqn:B.14}
{\cal T}_{10,T}^{(1)}&=& C_F \Bigg [{\sr \over 16}(4 + \rho) F_1 -{5 \rho
\over 8} F_2+ {\sr \over 4}(-20 + 7 \rho) F_3 - {1 \over 2} \sr \beta F_4
-{3 \over 4} \sr \beta
\nonumber \\[2ex]
&& + {\sr \over 16} (-156 + 57 \rho) {\cal L}i_2(t) +{\sr \over 8} (26
-13 \rho) \ln(t)
\nonumber \\[2ex]
&&+ {\sr \over 4}(-12 +3 \rho) \ln(t) \ln(1+t) \Bigg]\,,
\\[2ex]
\label{eqn:B.15}
{\cal T}_{11,T}^{(1)}&=& C_F \Bigg [ {\sr \over 4} F_1 + {\rho \over 8} (-5 
+\rho) F_2 -\sr F_3 +{1 \over 8} \sr \beta (14-3 \rho)
\nonumber \\[2ex]
&&-{7 \sr \over 4} {\cal L}i_2(t)+{\sr \over 16} (20 -12 \rho +3 \rho^2) \ln(t)
-\sr \ln(t) \ln(1+t) \Bigg]\,.
\end{eqnarray}
The normal polarized structure functions are
\begin{eqnarray}
\label{eqn:B.16}
{\cal T}_{9,N}^{(1)}&=& C_F \Bigg [-{\pi \over 4}\sr (1-\rho) \Bigg]\,,
\\[2ex]
\label{eqn:B.17}
{\cal T}_{12,N}^{(1)}&=& C_F \Bigg [- {\pi \over 4}\sr \beta (1+\rho) \Bigg]\,,
\\[2ex]
\label{eqn:B.18}
{\rm Im} {\cal T}_{13,N}^{(1)}&=& C_F \Bigg [ \sr \beta (\rho -2) G_1
          +{\sr \over 4} (8-13 \rho) G_2 +{\sr \over 8} (20+9 \rho)
          -{\rho \over 8} (3 \rho +26) 
\nonumber  \\[2ex]
&&-2 \sr (1-\rho) \ln(1+t) +{\sr \over 16} (24-2 \rho -3 \rho^2) 
   \ln(1+t-\sqt)
\nonumber  \\[2ex]
&&-4 \sr (1-\rho) \ln(1-\sqt)
+{\sr \over 32} (40-62 \rho -8 \beta +48 \rho \beta +3 \rho^2)\ln(t) \Bigg]\,,
\nonumber\\[2ex]
\end{eqnarray}
where $\Li(x)$ is defined in \cite{lbmr}. Furthermore the functions $F_i$ 
($i=1-4$) and $G_i$ ($i=1,2$) appearing in the expressions above are given by
\begin{eqnarray}
\label{eqn:B.19}
F_1&=& \Li(t^3) +4 \zeta(2) +{1 \over 2} \ln^2(t)+3 \ln(t) \ln(1+t+t^2)\,,
\nonumber\\[2ex]
F_2&=& \Li\left(-t^{3/2}\right) - \Li\left(t^{3/2}\right)
+ \Li\left(-t^{1/2}\right) - \Li\left(t^{1/2}\right) +3 \zeta(2)
\nonumber\\[2ex]
&& +2 \ln(t) \ln(1+\sqt) -2 \ln(t) \ln(1-\sqt)+{3 \over 2} \ln(t) \ln(1+t-\sqt)
\nonumber\\[2ex]
&&     -{3 \over 2} \ln(t) \ln(1+t+\sqt)\,,
\nonumber\\[2ex]
F_3&=&\Li(-t) + \ln(t) \ln(1-t)\,,
\nonumber\\[2ex]
F_4&=&6 \ln(t) -8 \ln(1-t) -4 \ln(1+t)\,,
\nonumber\\[2ex]
G_1&=& \Li\left(-t^{3/2}\right) - 3\Li\left(-t^{1/2}\right)
       - 4\Li\left(t^{1/2}\right) - \Li\left(-t\right)
\nonumber\\[2ex]
&&-{1 \over 2} \zeta(2) -{1 \over 8} \ln^2(t)\,,
\nonumber\\[2ex]
G_2&=& \Li\left({\sqt \over 1+t}\right)-{1 \over 8} \ln^2(t)
 -{1 \over 2} \ln(t) \ln(1+t) + {1 \over 2} \ln^2(1+t) -{1 \over 2} \zeta(2)\,,
\end{eqnarray}
where the variable t is defined in  (\ref{eqn:3.12}). Further we are interested
in the values taken by ${\cal T}_i$ when $m \rightarrow 0$. In the unpolarized
case they become
\begin{eqnarray}
\label{eqn:B.22}
{\cal T}_1^{(1),m \rightarrow 0}&=&{\cal T}_2^{(1),m \rightarrow 0}=1 \,,
\qquad {\cal T}_3^{(1),m \rightarrow 0}={\cal T}_4^{(1),m \rightarrow 0} =2\,, 
\nonumber\\[2ex]
&& {\cal T}_5^{(1),m \rightarrow 0}=0i\,,
\qquad {\cal T}_6^{(1),m \rightarrow 0}=0\,,
\end{eqnarray}
whereas the longitudinal structure functions tend to the limits
\begin{eqnarray}
\label{eqn:B.20}
{\cal T}_{7,L_1}^{(1),m \rightarrow 0}&=&{1 \over 4} \,, 
\qquad {\cal T}_{7,L_2}^{(1),m \rightarrow 0}=-1\,,
\qquad {\cal T}_{8,L}^{(1),m \rightarrow 0}=0\,,
\nonumber\\[2ex]
&& {\cal T}_{10,L}^{(1),m \rightarrow 0}
={\cal T}_{11,L}^{(1),m \rightarrow 0}= {1 \over 2}\,.
\end{eqnarray}
All transverse and normal polarized quantities become zero in the limit
$m \rightarrow 0$.

%


\centerline{\bf \large{FIGURE CAPTIONS}}
\begin{description}
\item[FIG. 1]
Longitudinal polarization $P_L$ of the up-quark up to second order in 
$\alpha_s$ at $Q=M_Z$ for polarized ($\lambda_{e^-}=\pm 1$) and unpolarized 
 (unpol) electrons. The positron beam is unpolarized. (a) Born $P_{L,0}$
(dashed line); (b) order $\alpha_s$ corrected $P_{L,1}$ (dashed dotted line); 
(c) order $\alpha_s^2$ corrected $P_{L,2}$
(solid line).
\item[FIG. 2]
Ratios $R_L$ of the higher order and lower order corrected longitudinal 
polarization of the up-quark at $Q=M_Z$ for unpolarized electrons and
positrons.
(a) Ratio of first order corrected and the Born contribution to the 
polarization $P_{L,1}/P_{L,0}$ (dashed line);
(b) Ratio of the second order corrected and the first order corrected 
polarization $P_{L,1}/P_{L,0}$ (solid line).
\item[FIG. 3]
Longitudinal polarization $P_L$ of the bottom-quark up to first order in 
$\alpha_s$ at $Q=M_Z$ for polarized ($\lambda_{e^-}=\pm 1$) and unpolarized 
(unpol) electrons. The positron beam is unpolarized. (a) Born $P_{L,0}$
(dashed line); (b) order $\alpha_s$ corrected $P_{L,1}$ (solid line).
\item[FIG. 4]
Ratios $R_L$ of order $\alpha_s$ contributions to various quantities
for unpolarized electrons and positrons for bottom production at $Q=M_Z$.
(a) Ratio of the longitudinal polarization in the massless quark
approach with $m_b=0$ and the exact longitudinal polarization $m_b=4.5$
(dotted line); (b)  Ratio of the unpolarized cross section $d^2\sigma/d~\Omega$
for $m_b\rightarrow 0$ and $m_b=4.5$ (dashed dotted line); (c)  Ratio of the 
longitudinal polarization in the massive quark approach with 
$m_b\rightarrow 0$ and the exact longitudinal polarization for $m_b=4.5$
(dashed line).
\item[FIG. 5]
Same as in Fig. 3 but now for top-quark production at $Q=500~{\rm GeV}$.
\item[FIG. 6]
Transverse polarization $P_T$ of the bottom-quark up to first order in
$\alpha_s$ at $Q=M_Z$ for polarized ($\lambda_{e^-}=\pm 1$) and unpolarized
(unpol) electrons. The positron beam is unpolarized. (a) Born $P_{T,0}$
(dashed line); (b) order $\alpha_s$ corrected $P_{T,1}$ (solid line).
\item[FIG. 7]
The same as in Fig. 6 but now top-quark production at $Q=500~{\rm GeV}$.
\item[FIG. 8]
Normal polarization $P_N$ of the bottom-quark up to first order in
$\alpha_s$ at $Q=M_Z$ for polarized ($\lambda_{e^-}=\pm 1$) and unpolarized
(unpol) electrons. The positron beam is unpolarized. (a) Born approximation
($\gamma-Z$-interference term) $P_{N,0}$ (dashed line); (b) order $\alpha_s$ 
(QCD) corrected results $P_{N,1}$ (solid line).
\item[FIG. 9]
Normal polarization $P_N$ of the top-quark up to first order in
$\alpha_s$ at $Q=500~{\rm GeV}$ for polarized ($\lambda_{e^-}=\pm 1$) and 
unpolarized (unpol) electrons. The positron beam is unpolarized. 
(a) $\lambda_{e^-}=-1$ (dashed line); (b) unpolarized electrons  (solid line);
(c) $\lambda_{e^-}=1$ (dotted line). The lower curves represent the Born
approximation ($\gamma-Z$-interference term) $P_{N,0}$ whereas the upper curves
stand for the order $\alpha_s$ (QCD) corrected results $P_{N,1}$.
\end{description}
\newpage

\vspace*{\fill}
\vbox{ \centerline{ \epsfig{file=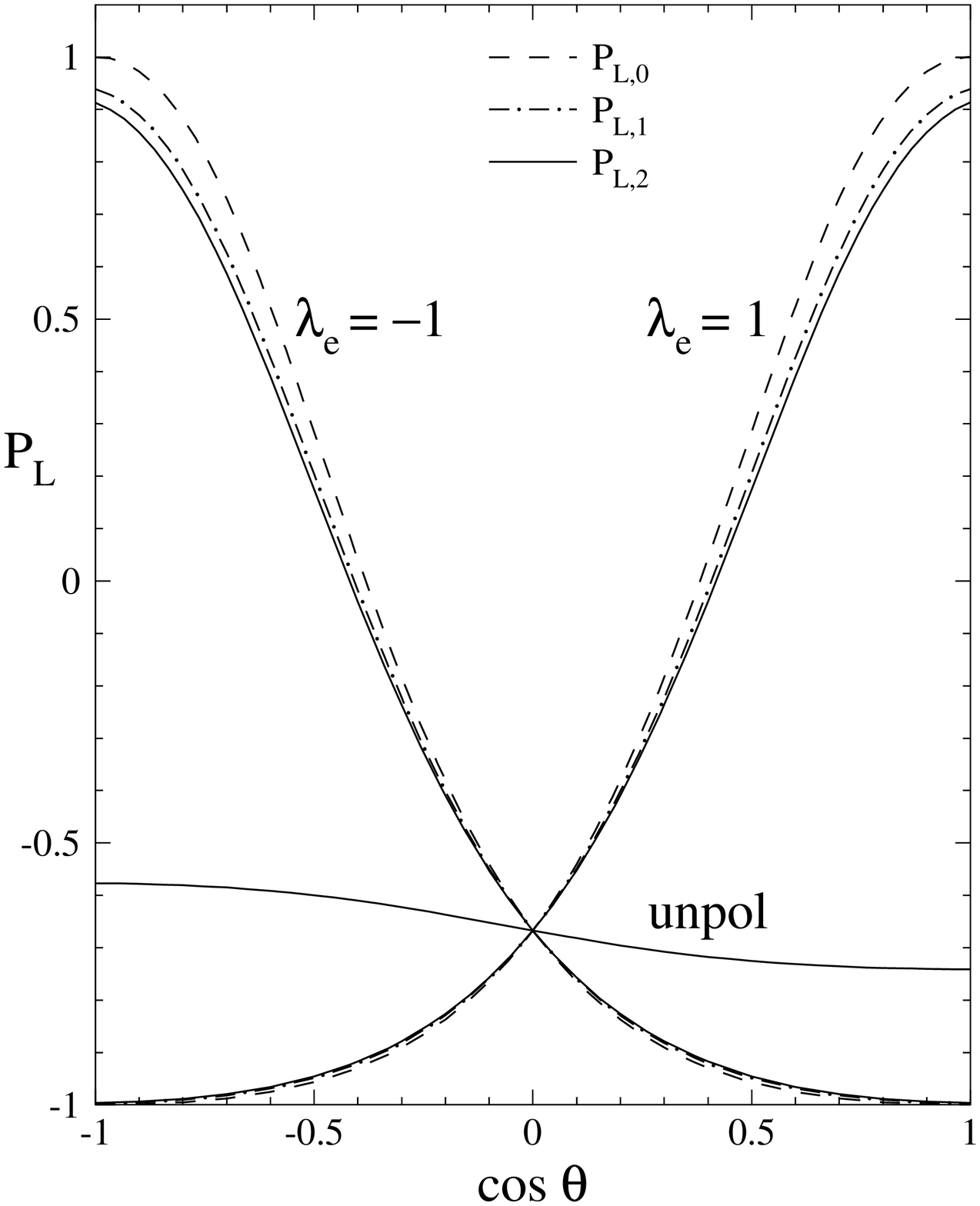,width=16cm,angle=0} } } 
\vspace*{\fill}
 
\vspace*{\fill}
\begin{center}
{\bf \Large{Fig.~1}}
\end{center}

\newpage

\vspace*{\fill}
\vbox{ \centerline{ \epsfig{file=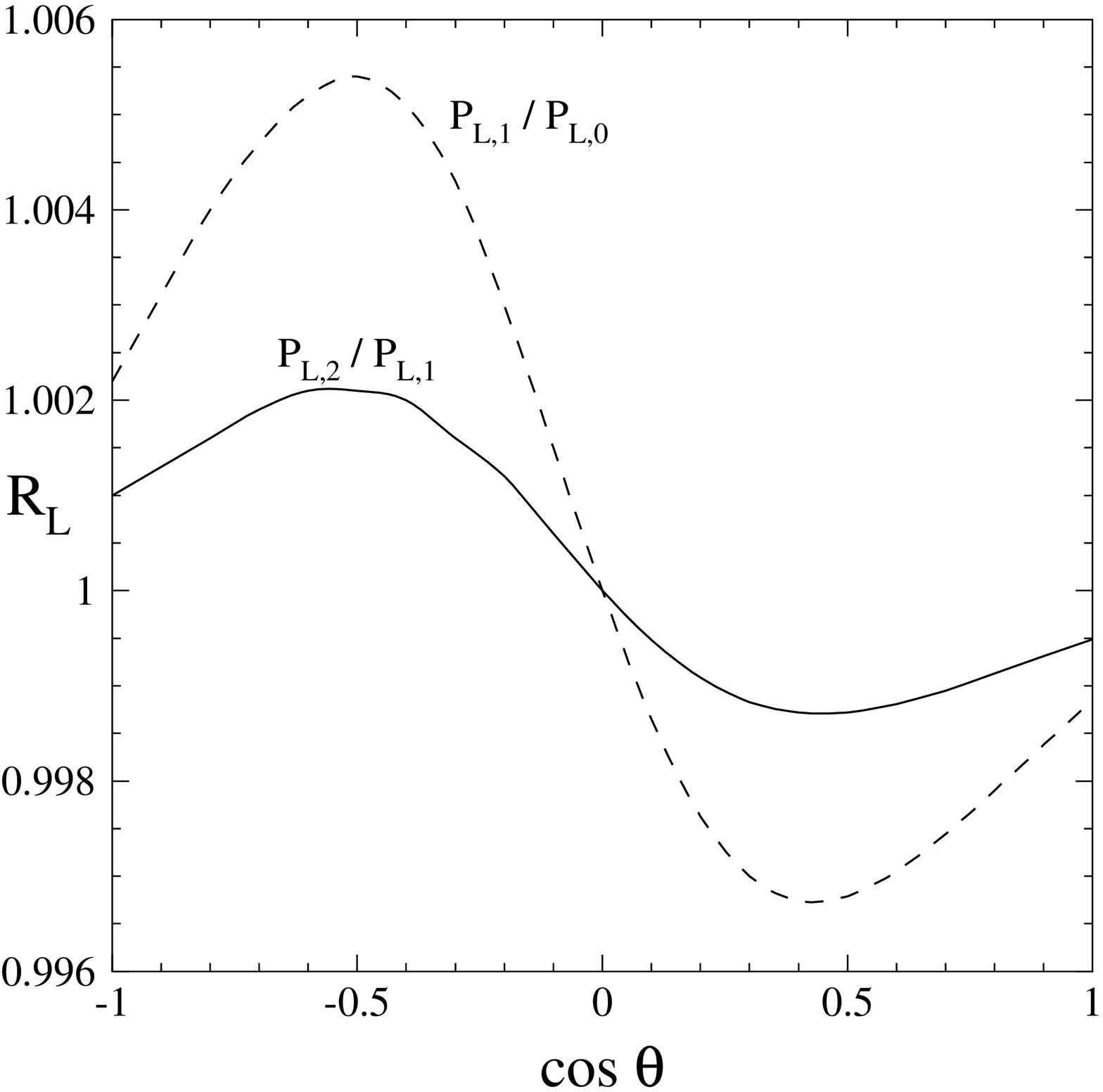,width=16cm,angle=0} } }
\vspace*{\fill}

\vspace*{\fill}
\begin{center}
{\bf \Large{Fig.~2}}
\end{center}

\newpage

\vspace*{\fill}
\vbox{ \centerline{ \epsfig{file=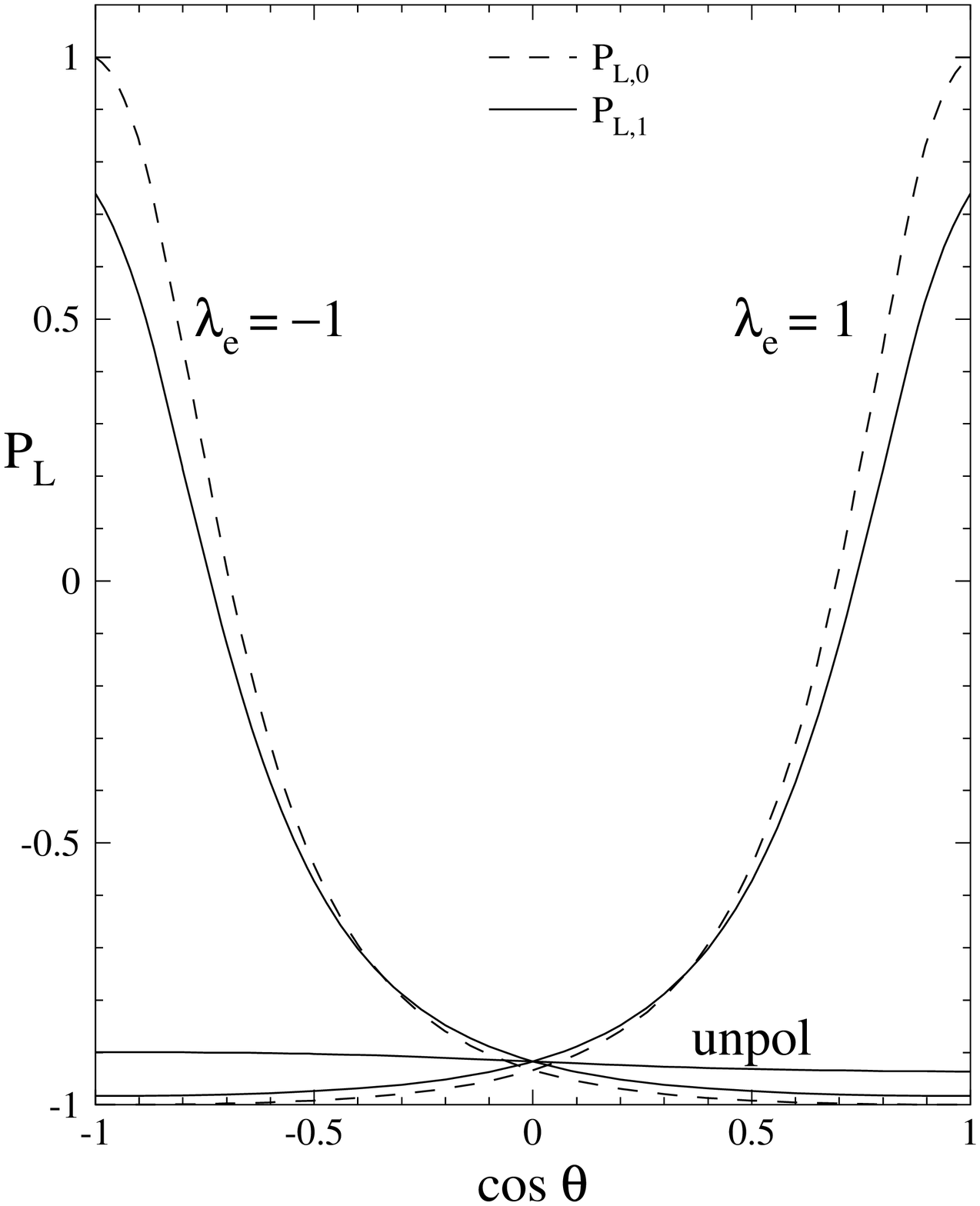,width=16cm,angle=0} } }
\vspace*{\fill}

\vspace*{\fill}
\begin{center}
{\bf \Large{Fig.~3}}
\end{center}

\newpage

\vspace*{\fill}
\vbox{ \centerline{ \epsfig{file=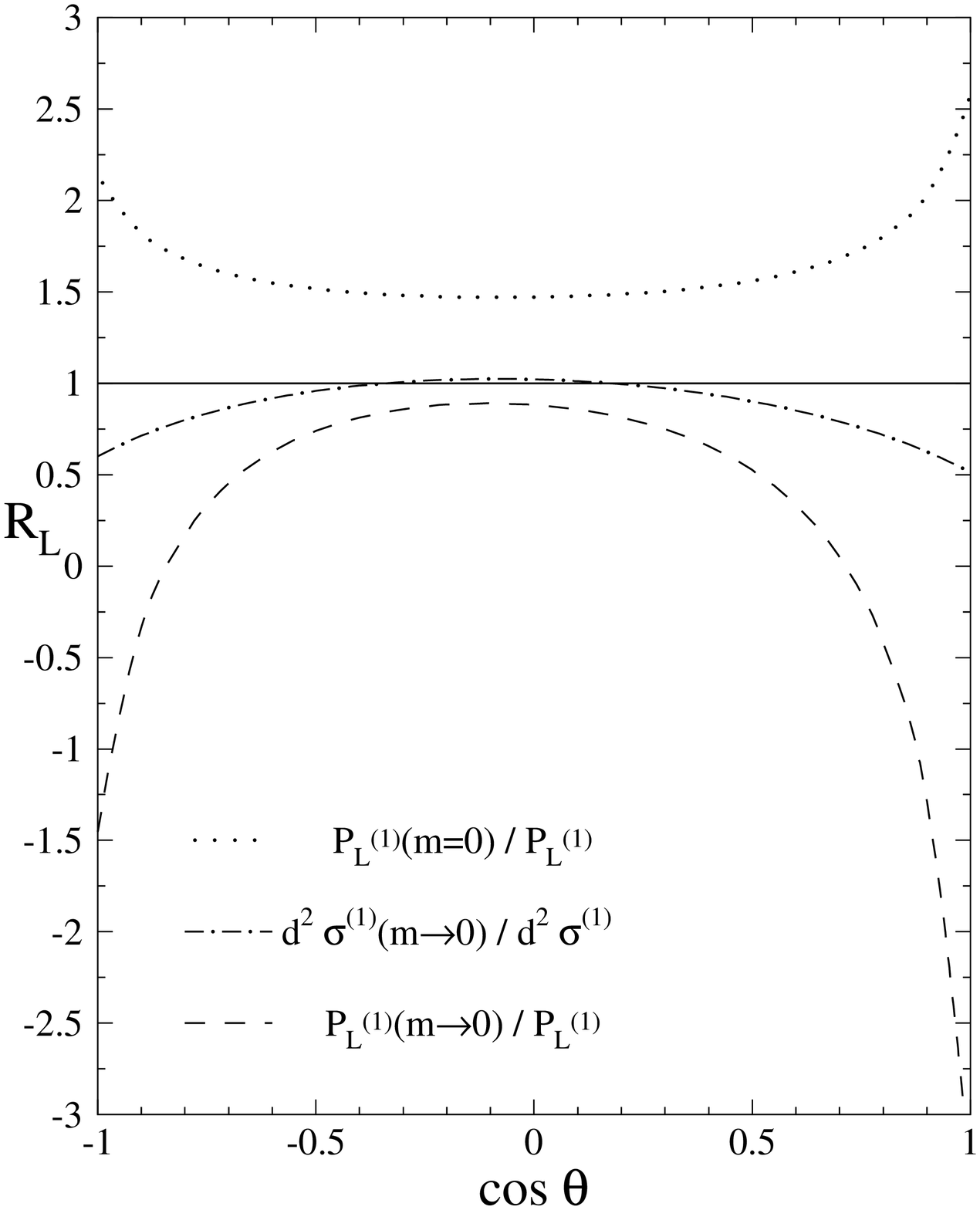,width=16cm,angle=0} } }
\vspace*{\fill}

\vspace*{\fill}
\begin{center}
{\bf \Large{Fig.~4}}
\end{center}

\newpage

\vspace*{\fill}
\vbox{ \centerline{ \epsfig{file=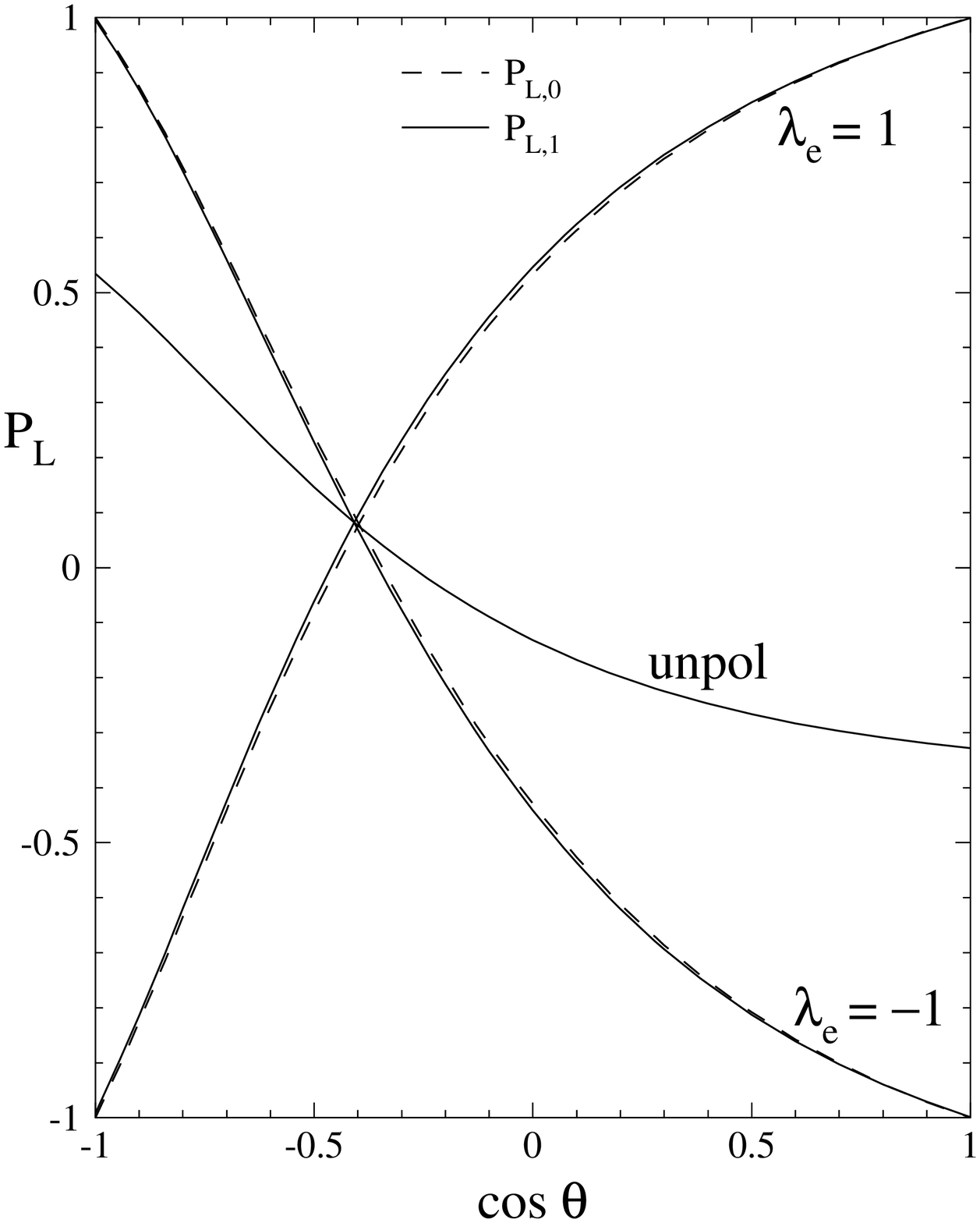,width=16cm,angle=0} } }
\vspace*{\fill}

\vspace*{\fill}
\begin{center}
{\bf \Large{Fig.~5}}
\end{center}

\newpage

\vspace*{\fill}
\vbox{ \centerline{ \epsfig{file=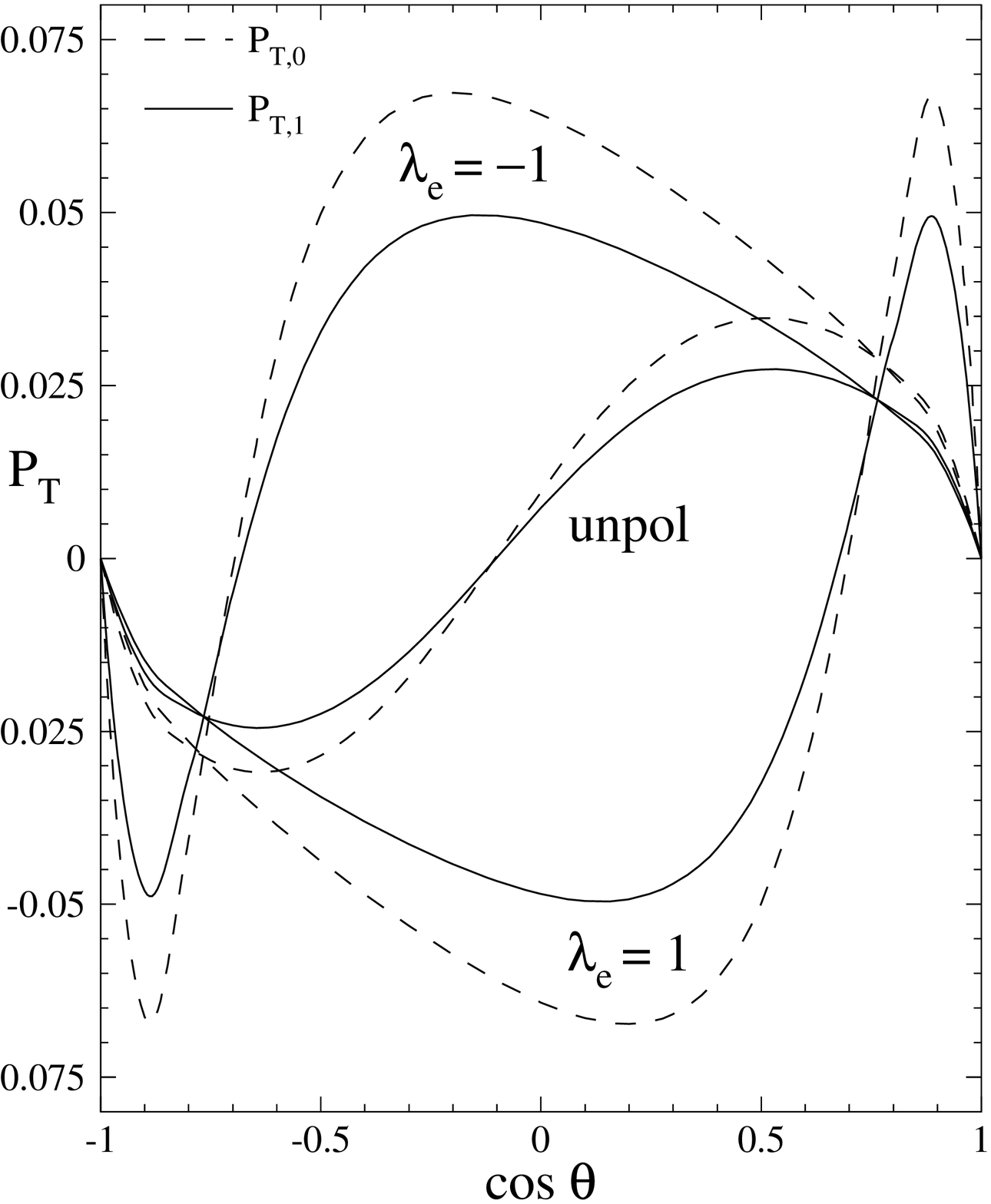,width=16cm,angle=0} } }
\vspace*{\fill}

\vspace*{\fill}
\begin{center}
{\bf \Large{Fig.~6}}
\end{center}

\newpage

\vspace*{\fill}
\vbox{ \centerline{ \epsfig{file=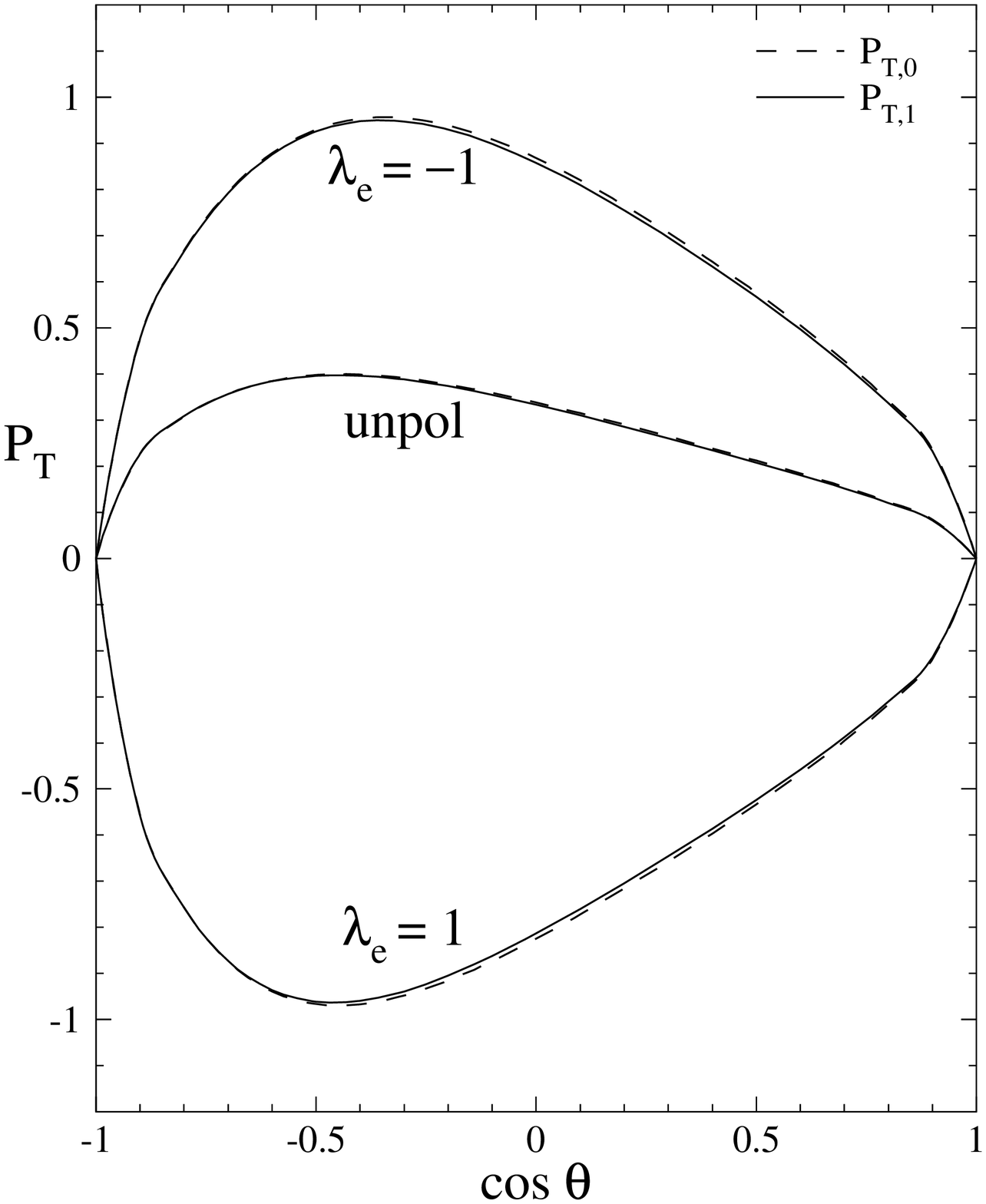,width=16cm,angle=0} } }
\vspace*{\fill}

\vspace*{\fill}
\begin{center}
{\bf \Large{Fig.~7}}
\end{center}

\newpage

\vspace*{\fill}
\vbox{ \centerline{ \epsfig{file=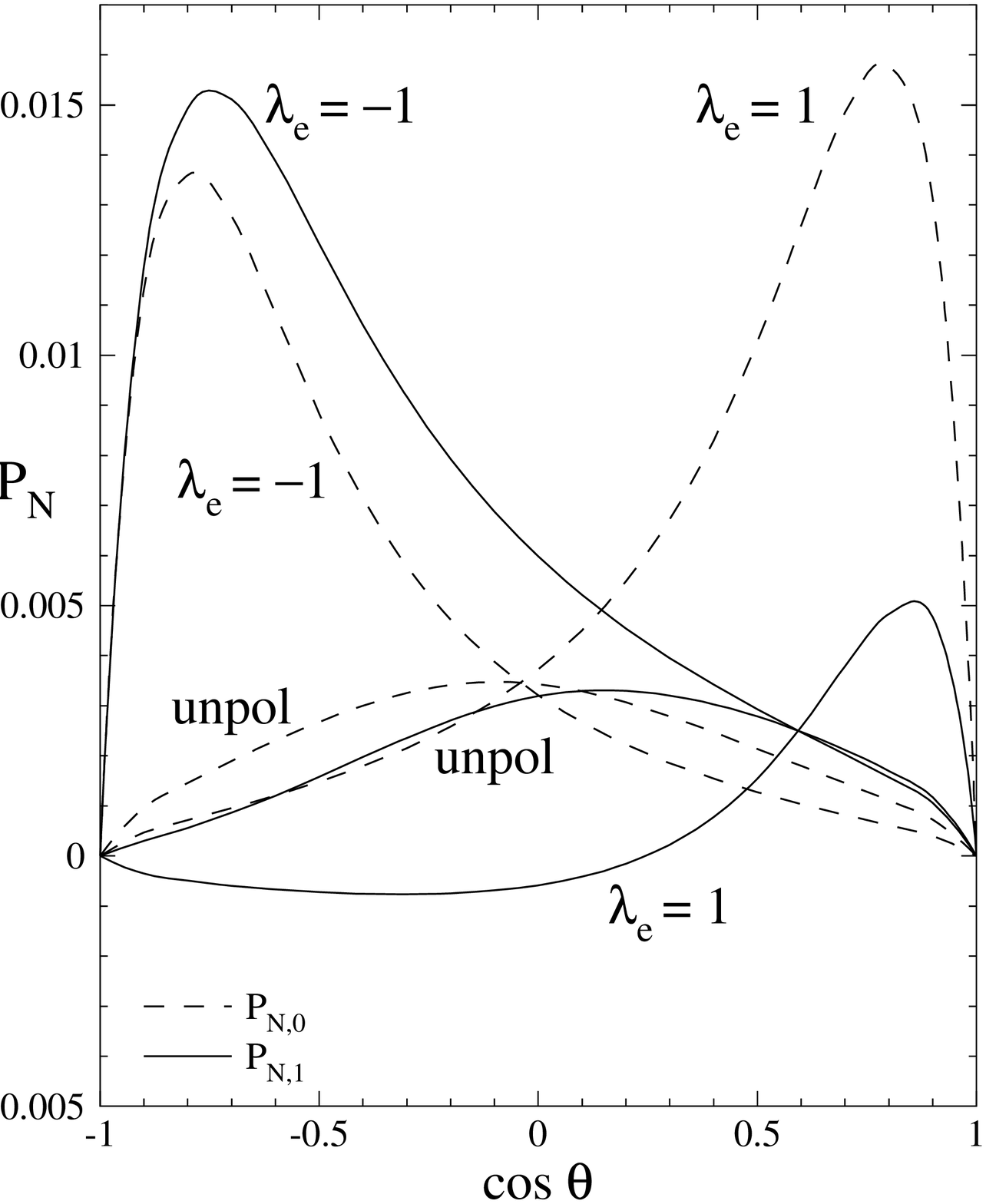,width=16cm,angle=0} } }
\vspace*{\fill}

\vspace*{\fill}
\begin{center}
{\bf \Large{Fig.~8}}
\end{center}

\newpage

\vspace*{\fill}
\vbox{ \centerline{ \epsfig{file=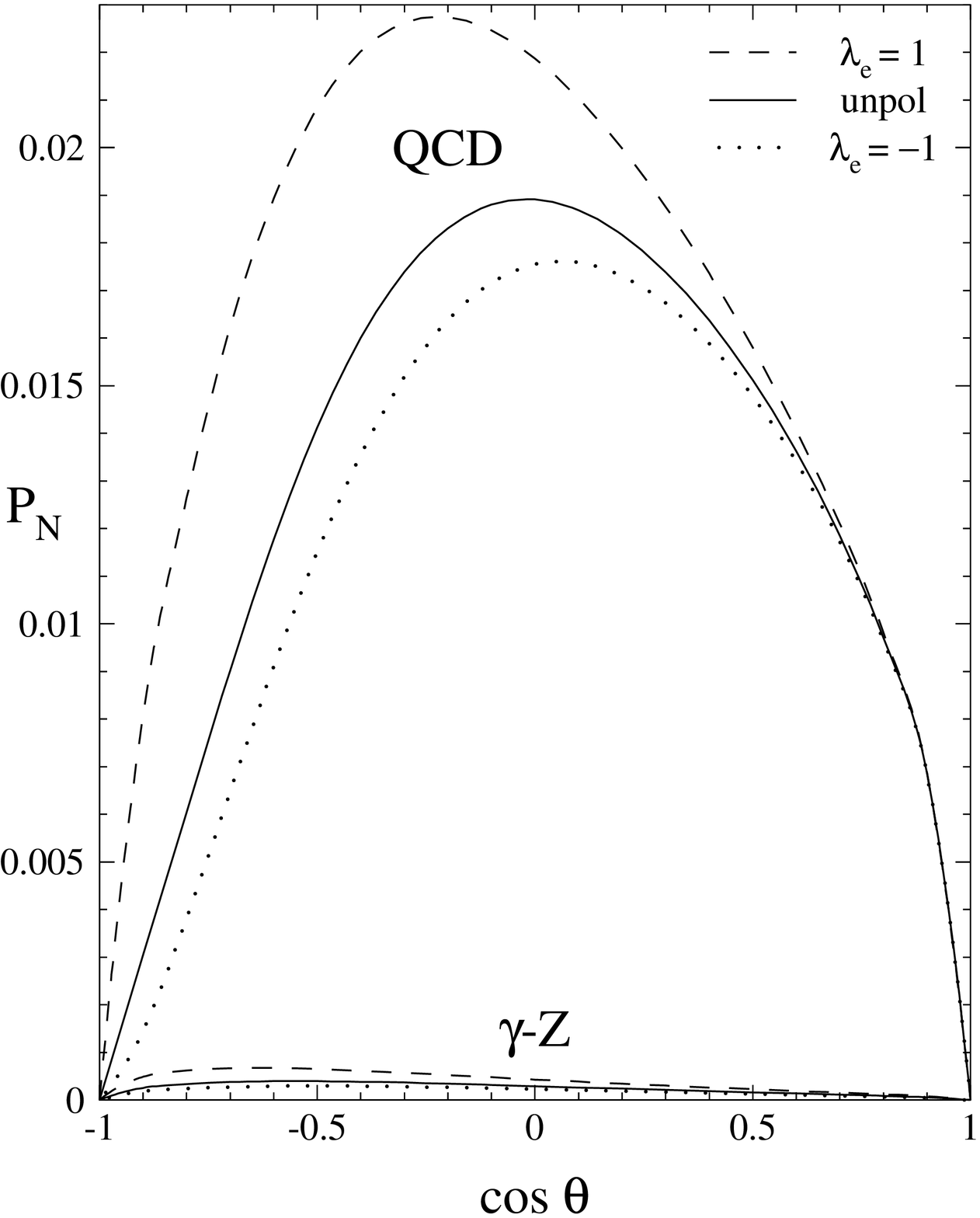,width=16cm,angle=0} } }
\vspace*{\fill}

\vspace*{\fill}
\begin{center}
{\bf \Large{Fig.~9}}
\end{center}



\begin{thebibliography}{99}
\bibitem{zerw}
Proceedings: e+ e- Collisions at TeV Energies: The Physics Potential: 
Edited by P.M. Zerwas. Hamburg, Germany, DESY, 1996. 550p. (DESY 96-123D)
\bibitem{ritu}
S.D. Rindani, M.M. Tung, Phys. Lett. {\bf B424} (1998) 125.
\bibitem{craig}
N.S. Craigie, Phys.Rept. {\bf 99} (1983) 69.
\bibitem{krz}
J.H. K\"uhn, A. Reiter and P.M. Zerwas, Nucl. Phys. {\bf B272} (1986) 560.
\bibitem{akp}
M. Anselmino, P. Kroll and B. Pire, Phys. Lett. {\bf B167} (1986) 113.
\bibitem{pash}
S. Parke and Y. Shadmi, Phys. Lett. {\bf B387} (1996) 199.
\bibitem{stol}
J.B. Stav and H. A. Olsen, Z. Phys. {\bf C57} (1993) 519, Phys. Rev.
{\bf D50} (1994) 6775, ibid. {\bf D52} (1995) 1359, {\bf D54} (1996) 817,
{\bf D56} (1997) 407.
\bibitem{kopitu}
J.G. K\"orner, A. Pilaftsis and M.M. Tung, Z. Phys. {\bf C63} (1994) 509.
\bibitem{grkotu1}
S. Groote, J.G. K\"orner, M.M. Tung, Z. Phys. {\bf C70} (1996) 281.
\bibitem{grko}
S. Groote, J.G. K\"orner, Z. Phys. {\bf C72} (1996) 255.
\bibitem{grkotu2}
S. Groote, J.G. K\"orner, M.M. Tung, Z. Phys. {\bf C74} (1997) 615.
\bibitem{konapa}
J. Kodaira, T. Nasuno and S. Parke, Phys. Rev. {\bf D59} (1999) 014023.
\bibitem{schm}
C. Schmidt, Phys. Rev. {\bf D54} (1996) 3250.
\bibitem{brfl}
A. Brandenburg, M. Flesch, and P. Uwer, Phys. Rev. {\bf D59} (1999) 014001,
Chech. J. of Phys. {\bf 50} (2000) Suppl. S1, 51.
\bibitem{rane1}
V. Ravindran and W.L. van Neerven, Phys.Lett. {\bf B445} (1998) 214.
\bibitem{rane2}
V. Ravindran and W.L. van Neerven, Phys.Lett. {\bf B445} (1998) 206.
\bibitem{lbmr}
L. Lewin, "Polylogarithms and Associated Functions", North Holland,
Amsterdam, 1983;\\
R. Barbieri, J.A. Mignaco and E. Remiddi, Nuovo Cimento {\bf 11A}
(1972) 824;\\
A. Devoto and D.W. Duke, Riv. Nuovo. Cimento Vol. 7, N. 6 (1984) 1.
\bibitem{rine1}
P.J. Rijken and W.L. van Neerven, Phys.Lett. {\bf B386} (1996) 422,
ibid. {\bf B392} (1997) 207, Nucl. Phys. {\bf B487} (1998) 233.
\bibitem{ne1}
W.L. van Neerven, Acta Phys.Polon. {\bf B29} (1998) 2573.
\bibitem{rine2}
P.J. Rijken and W.L. van Neerven, Nucl. Phys. {\bf B523} (1998) 245.
\bibitem{case}
S. Catani and M.H. Seymour, JHEP 9907:023, (1999).
\bibitem{fase}
B. Falk and L.M. Sehgal, Phys. Lett. {\bf B325} (1994) 509.
\bibitem{flsa}
D. de Florian and R. Sassot, Nucl. Phys. {\bf B488} (1997) 367.
\bibitem{ra}
V. Ravindran, Phys. Lett. {\bf B398} (1997) 169, Nucl. Phys. {\bf B490} 
(1997) 272.
\bibitem{stvo}
M. Stratmann and W. Vogelsang, Nucl. Phys. {\bf B496} (1997) 41.
\bibitem{mene}
R. Mertig and W.L. van Neerven, Z. Phys. {\bf C60} (1993) 489, Erratum
ibid. {\bf C65} (1995) 360.
\bibitem{teve}
O.V. Teryaev and O.L. Veretin, preprint hep-ph/9602362.
\bibitem{chkust}
K.G. Chetyrkin, J.H. K\"uhn and M. Steinhauser, Phys. Lett. {\bf B371} (1996)
93, Nucl. Phys. {\bf B482} (1996) 213.
\bibitem{ne2}
W.L. van Neerven, Acta Phys.Polon. {\bf B29} (1998) 1175.
\bibitem{caso}
C. Caso et al., Review of Particle Physics, Eur. Phys. J. {\bf C3} (1998) 19, 
24, 87.
\bibitem{hage}
R. Hagedorn, Relativistic Kinematics, W.A. Benjamin, Inc., New York and
Amsterdam, 1964.
\end{thebibliography}
\end{document}